\documentclass[12pt]{article}\usepackage[]{graphicx}\usepackage[]{color}
\makeatletter
\def\maxwidth{ %
  \ifdim\Gin@nat@width>\linewidth
    \linewidth
  \else
    \Gin@nat@width
  \fi
}
\makeatother

\definecolor{fgcolor}{rgb}{0.345, 0.345, 0.345}
\newcommand{\hlnum}[1]{\textcolor[rgb]{0.686,0.059,0.569}{#1}}%
\newcommand{\hlstr}[1]{\textcolor[rgb]{0.192,0.494,0.8}{#1}}%
\newcommand{\hlcom}[1]{\textcolor[rgb]{0.678,0.584,0.686}{\textit{#1}}}%
\newcommand{\hlopt}[1]{\textcolor[rgb]{0,0,0}{#1}}%
\newcommand{\hlstd}[1]{\textcolor[rgb]{0.345,0.345,0.345}{#1}}%
\newcommand{\hlkwb}[1]{\textcolor[rgb]{0.69,0.353,0.396}{#1}}%
\newcommand{\hlkwc}[1]{\textcolor[rgb]{0.333,0.667,0.333}{#1}}%
\newcommand{\hlkwd}[1]{\textcolor[rgb]{0.737,0.353,0.396}{\textbf{#1}}}%

\usepackage{framed}
\makeatletter
\newenvironment{kframe}{%
 \def\at@end@of@kframe{}%
 \ifinner\ifhmode%
  \def\at@end@of@kframe{\end{minipage}}%
  \begin{minipage}{\columnwidth}%
 \fi\fi%
 \def\FrameCommand##1{\hskip\@totalleftmargin \hskip-\fboxsep
 \colorbox{shadecolor}{##1}\hskip-\fboxsep
     \hskip-\linewidth \hskip-\@totalleftmargin \hskip\columnwidth}%
 \MakeFramed {\advance\hsize-\width
   \@totalleftmargin\z@ \linewidth\hsize
   \@setminipage}}%
 {\par\unskip\endMakeFramed%
 \at@end@of@kframe}
\makeatother

\definecolor{shadecolor}{rgb}{.97, .97, .97}
\definecolor{messagecolor}{rgb}{0, 0, 0}
\definecolor{warningcolor}{rgb}{1, 0, 1}
\definecolor{errorcolor}{rgb}{1, 0, 0}
\newenvironment{knitrout}{}{} 

\usepackage{alltt}
\usepackage[margin=1in]{geometry}
\usepackage{graphicx}
\usepackage[affil-it]{authblk}
\usepackage{natbib}
\usepackage[bf,font={small,sl}]{caption}

\usepackage{amsmath}
\usepackage{amsfonts}
\usepackage{bm}
\usepackage{url}
\usepackage{graphicx}

\usepackage{lineno}

\title{\textbf{Varying-coefficient stochastic differential equations with applications in ecology}}

\author{Th\'eo Michelot\footnote{Email: tm75@st-andrews.ac.uk}, Richard Glennie, Catriona Harris, Len Thomas}
\affil{University of St Andrews, UK}
\date{}
\IfFileExists{upquote.sty}{\usepackage{upquote}}{}
\begin{document}

\maketitle

\begin{abstract}
Stochastic differential equations (SDEs) are popular tools to analyse time series data in many areas, such as mathematical finance, physics, and biology. They provide a mechanistic description of the phenomeon of interest, and their parameters often have a clear interpretation. These advantages come at the cost of requiring a relatively simple model specification. We propose a flexible model for SDEs with time-varying dynamics where the parameters of the process are non-parametric functions of covariates, similar to generalized additive models. Combining the SDE and non-parametric approaches allows for the SDE to capture more detailed, non-stationary, features of the data-generating process. We present a computationally efficient method of approximate inference, where the SDE parameters can vary according to fixed covariate effects, random effects, or basis-penalty smoothing splines. We demonstrate the versatility and utility of this approach with three applications in ecology, where there is often a modelling trade-off between interpretability and flexibility.    
\end{abstract}

\newpage
\section{Introduction}

Stochastic differential equations (SDEs) describe the evolution of a system that involves stochastic noise \citep{allen2007}. We present a general approach to improving the flexibility of such models. To introduce it, we focus on the most popular form of SDE,
\begin{linenomath*}
\begin{equation}
  \label{eq:SDE}
  \mathop{dZ_t} = \mu(Z_t, t) \mathop{dt} + \sigma(Z_t, t) \mathop{dW_t},\quad Z_0 = z_0,
\end{equation}
\end{linenomath*}
where $(W_t)$ is a Wiener process, and $z_0$ is a known initial condition. The terms of the equation describe the evolution of the process $(Z_t)$: the drift $\mu$ measures the expected change in the process over an infinitesimal time interval, and the diffusion $\sigma$ captures variability. In most applications, the drift and diffusion are chosen as simple parametric functions; the objective is to estimate their parameters and obtain a mechanistic description of the system. SDEs have, e.g., been applied in finance to study asset pricing \citep{ait2007}, in biology to describe population dynamics \citep{dennis1991}, and in epidemiology to predict disease spread \citep{allen2006}. Eq.\ \ref{eq:SDE} includes Brownian motion, geometric Brownian motion, and the Ornstein-Uhlenbeck process as special cases.

SDEs are used to formulate a (simplified) description of a stochastic system, and the challenge is to build models flexible enough to reflect features of the system within the assumed structure of Eq.\ \ref{eq:SDE}. For this purpose, there has been interest in specifying SDEs with time-varying dynamics. Regime-switching models have been developed, where a process switches between a finite number of SDEs, often based on an underlying continuous-time Markov chain \citep{mao2006}. These models combine the convenience of simple parametric models with the flexibility provided by multiple regimes, and have been used to describe, e.g.,  the movement of animals switching between behavioural states \citep{blackwell1997} or the time-varying dynamics of oil prices \citep{liechty2001}. An alternative is to specify the parameters of a SDE as continuous-valued random processes \citep[e.g.][]{duan2009}; e.g., in stochastic volatility models, the variance parameter of the diffusion function is itself specified as a diffusion process to account for changes in the variability \citep{ait2007}. Other approaches have been developed to model the drift and diffusion of SDEs as non-parametric functions of time or of the value of the process $Z_t$, e.g.\ using Gaussian processes \citep{archambeau2007} or orthogonal Legendre polynomials \citep{rajabzadeh2016}. For a particular class of SDEs applied to animal movement studies, \cite{preisler2004} and \cite{russell2018} suggested using splines to model the dependence of SDE parameters on spatial covariates.

We propose a general approach where the parameters of a SDE are specified as basis-penalty smoothing splines, similar to generalized additive models \citep[GAMs;][]{wood2017}. This allows for a rich class of models including linear covariate effects, factor variables, independent random effects, and smooth (non-parametric) covariate effects. It stands in contrast to the regime-switching models where parameters are piecewise constant rather than smooth. It generalises the models where parameters are specified as Gaussian processes, given the equivalent interpretation of Gaussian processes and smoothing splines \citep[][Section 5.8.2]{wood2017}, and it also extends ideas from \cite{preisler2004} and \cite{russell2018} to allow for flexible covariate dependence in a more general class of SDEs. We develop a method of inference, using a smoothing penalty in the SDE likelihood to control the roughness of non-parametric terms. We present a computationally efficient implementation based on the R packages mgcv and TMB, for model specification and model fitting, respectively.

We illustrate the potential of this new framework using three case studies from ecology. SDEs have great theoretical and practical appeal for the analysis of ecological data, because their continuous-time formulation does not depend on the sampling resolution of the data. Inferences from these models can therefore be compared across studies with different sampling schemes, and they can be fitted to data collected at irregular time intervals \citep[e.g.][]{michelot2019ctcrw}. Despite their advantages, continuous-time models have been underutilised in this field, in part because they have lacked flexibility to specify time-varying dynamics and covariate effects, or have required computationally-costly model fitting procedures. The three case studies illustrate the utility of the new model over existing parametric approaches and highlight its flexibility and computational convenience. In the supplementary materials, we give implementation details, a simulation study, another case study from finance, and the source code to reproduce all analyses presented in the paper. The method that we describe is implemented in an R package, available at \url{github.com/TheoMichelot/smoothSDE}.

\section{Varying-coefficient stochastic differential equations}
\label{sec:model}

\subsection{Model formulation}
\label{sec:model_general}

We consider a stochastic process $(Z_t)$ defined by 
\begin{linenomath*}
\begin{equation}
    \label{eq:SDE2}
    dZ_t = \mu(Z_t, \bm\theta_t)\ dt + \sigma(Z_t, \bm\theta_t)\ dW_t,    
\end{equation}
\end{linenomath*}
where the drift $\mu$ and diffusion $\sigma$ depend on a time-varying parameter vector $\bm\theta_t$. We assume that $\mu$ and $\sigma$ are known functions of $Z_t$ and $\bm\theta_t$; they determine the type of stochastic process (e.g., Brownian motion, Ornstein-Uhlenbeck process). The parameter $\bm\theta_t$ depends on time through its relationship with $J$ temporal covariates $x_{1t}, x_{2t}, \dots, x_{Jt}$, and we write each component $\theta_t$ of $\bm\theta_t$ as
\begin{linenomath*}
\begin{equation*}
  h(\theta_t) = \beta_0 + f_1(x_{1t}) + f_2(x_{2t}) + \dots + f_J(x_{Jt}),
\end{equation*}
\end{linenomath*}
where $h$ is a link function, $\beta_0$ is an intercept parameter and, for $j = 1, \dots, J$, $f_j$ could be a linear effect of a covariate, an independent random effect, or a smooth function. A simple example would be to have $x_{1t} = t$, to express that the dynamics of the process depend on time. For smooth functions or random effects, we employ the basis-penalty approach \citep{wood2017}, writing the functions as linear combinations of $m_j$ basis functions $\{ \psi_{jk} \}$,
\begin{linenomath*}
\begin{equation}
  \label{eq:spline}
  f_j(x) = \sum_{k=1}^{m_j} \beta_{jk} \psi_{jk}(x),
\end{equation}
\end{linenomath*}
where several standard bases could be considered, e.g., cubic splines, thin plate regression splines, or B-splines. We will refer to this model as a varying-coefficient stochastic differential equation, as an analogy with the varying-coefficient models of \cite{hastie1993}.

\subsection{Model fitting}
\label{sec:model_fitting}

We consider $n$ observations $( z_1, z_2, \dots, z_n )$ from the process $(Z_t)$, collected at (possibly irregular) times $t_1 < t_2 < \dots < t_n $. The aim is to estimate the relationship between the parameters $\bm\theta_t$ governing the drift and diffusion of the process and the covariates. The method that we propose is based on (1) the likelihood of the observations under the SDE model, and (2) a penalty added to the likelihood to control the roughness of non-parametric terms in $\bm\theta_t$.

\subsubsection{Likelihood}
\label{sec:model_lk}

Diffusion processes are Markovian, so the likelihood of $n$ observations can be obtained as the product of the likelihoods of the individual transitions,
\begin{linenomath*}
\begin{align}
    \label{eq:lk}
    L(\bm\alpha, \bm\beta \vert z_1, \dots, z_n) & = [Z_{t_1} = z_1, \dots, Z_{t_n} = z_n] \nonumber \\
        & = [Z_{t_1} = z_1] \prod_{i=1}^{n-1} [Z_{t_{i+1}} = z_{i+1} \vert Z_{t_i} = z_i],  
\end{align}
\end{linenomath*}
where $\bm\beta$ contains the basis coefficients from Eq.\ \ref{eq:spline} and $\bm\alpha$ is the vector of other parameters of the model (e.g., linear covariate effects), and where $[\cdot]$ is the pdf. The dependence on $\bm\alpha$ and $\bm\beta$ is omitted in the right-hand side of Eq.\ \ref{eq:lk} for notational simplicity. We assume that the first value $z_1$ is deterministic, such that $[Z_{t_1} = z_1] = 1$. 

Evaluating the likelihood requires computation of the transition density $[Z_{t_{i+1}} \vert Z_{t_i}]$ of the process. For many common processes, such as Brownian motion, geometric Brownian motion, and the Ornstein-Uhlenbeck process, this density has an analytical expression. In such cases, the transition density of the corresponding varying-coefficient process can be approximated by assuming that the parameter $\bm\theta_t$ is fixed over each time interval of observation. Then, the time-varying parameter $\bm\theta_{t_i}$ can be substituted into the transition density of the standard process. However, many SDEs of the form given in Eq.\ \ref{eq:SDE2} do not have a closed-form transition density. More generally, we can then use the Euler-Maruyama discretization, and approximate the transition density $[Z_{t_{i+1}} \vert Z_{t_i} = z_i ]$ by the pdf of a normal distribution with mean $z_i + \mu(z_i, \bm\theta_{t_i}) \Delta_i$ and variance $\sigma(z_i, \bm\theta_{t_i})^2 \Delta_i$, where $\Delta_i = t_{i+1} - t_i$. This approximation assumes that the drift and diffusion terms are constant over each interval $[t_i, t_{i+1})$ between two observations. We present the varying-coefficient versions of several common processes in Appendix A, and give their approximate transition densities. Substituting the approximate transition density into Eq.\ \ref{eq:lk} yields the approximate likelihood for the full data set.

This method of inference is not exact, because it uses the transition density of the time-discretized diffusion process. The Euler-Maruyama discretization has the advantage of being widely applicable and easy to implement, but the accuracy of the estimation will decrease for longer time intervals between observations. To mitigate the effects of this approximation, we could include additional time points in the time series of observed data, corresponding to ``missing'' observations, and integrate over them, e.g., using either Markov chain Monte Carlo methods or the Laplace approximation \citep{elerian2001, albertsen2019}. Adding these missing values to the grid of observations leads to a finer time resolution, and improves the accuracy of the approximation, such that the error can be made arbitrarily small.

The process $(Z_t)$ might sometimes not be observed directly, in which case the problem of inference is slightly different. This can be viewed as a state-space model, where the state equation is given by the transition density $[Z_{t_{i+1}} \vert Z_{t_i}]$ (e.g., obtained using the Euler approximation), and the observation equation is the density $[\tilde{Z}_{t_i} \vert Z_{t_i} ]$, where $\tilde{Z}_{t_i}$ denotes the observations. In this context, $\tilde{Z}_t$ could e.g.\ include measurement error or be a more general function of $Z_t$. The diffusion process of interest is a latent process in the model, and it must be marginalised over to obtain the likelihood of the observed data. In the case of a Gaussian linear state-space model, the Kalman filter can be implemented, with time-varying parameters, and the likelihood obtained as a by-product. In this case, the Kalman filter can also be used to integrate over missing data in a data augmentation scheme to improve the discretization approximation. One example of a latent-state SDE is the velocity Ornstein-Uhlenbeck model described by \cite{johnson2008}, where the observed process (location) is the integral of a diffusion process (velocity). In that model, the location process is smooth, and it is therefore convenient to describe the persistent movement of an animal or particle, or other processes with strong autocorrelation. Non-Gaussian state-space models can also be accommodated using Markov chain Monte Carlo or the Laplace approximation to marginalise over the state process, as suggested, e.g., by \cite{albertsen2015}. We present two examples of state-space SDE models in Section \ref{sec:examples}.

\subsubsection{Smoothing penalty}
\label{sec:model_penalty}

Within the basis-penalty approach of GAMs, the roughness of the smoothing splines can be penalised in the likelihood, to obtain smooth relationships between the parameter $\bm\theta_t$ and the covariates. The penalised log-likelihood is
\begin{linenomath*}
\begin{equation}
  \label{eq:penlk}
  l_p(\bm\alpha, \bm\beta, \bm\lambda \vert z_1, \dots, z_n) = 
    \log\{L(\bm\alpha, \bm\beta \vert z_1, \dots, z_n)\} - \sum_j \lambda_j \bm\beta_j^T \bm{S}_j \bm\beta_j,
\end{equation}
\end{linenomath*}
where $L(\bm\alpha, \bm\beta \vert z_1, \dots, z_n)$ is the unpenalised likelihood given in Eq.\ \ref{eq:lk}, $\bm\beta_j$ is the vector of basis coefficients, $\bm{S}_j$ is the smoothing matrix associated with the chosen penalty, and $\lambda_j$ is a smoothness parameter for the $j$-th smooth term in $\bm\theta_t$ \citep{wahba1990}. $\bm{S}_j$ is a matrix of known coefficients, and it is constructed such that $\bm\beta_j^T \bm{S}_j \bm\beta_j$ measures the roughness (wiggliness) of the corresponding smooth term \citep{wood2017}. The penalised log-likelihood can then be used to perform maximum likelihood estimation, or Bayesian inference can be performed if the penalty is viewed as an improper prior on the basis coefficients. 

In Eq.\ \ref{eq:penlk}, the penalised log-likelihood is expressed in terms of the degrees of smoothness $\bm\lambda = (\lambda_{1}, \lambda_{2}, \dots)$ of the smoothing splines. In most applications, $\bm\lambda$ is unknown, and it must be estimated from the data. Here, we consider the marginal likelihood approach, i.e., we treat the basis coefficients $\bm\beta$ of the splines as random effects, and integrate them out of the likelihood. This yields the marginal likelihood of the smoothness parameters $\bm\lambda$ and other fixed parameters $\bm\alpha$,
\begin{linenomath*}
\begin{equation}
  \label{eq:marginal}
  L(\bm\alpha, \bm\lambda \vert z_1, \dots, z_n) = 
    \int L(\bm\alpha, \bm\beta \vert z_1, \dots, z_n) [\bm\beta \vert \bm\lambda] d\bm\beta,
\end{equation}
\end{linenomath*}
where $[\bm\beta \vert \bm\lambda]$ is the density of a multivariate normal distribution with mean zero and block-diagonal precision matrix. Each block of the precision matrix corresponds to the penalty for the basis coefficients of one smoothing spline, and it can be written $\lambda_j \bm{S}_j$. As with standard GAMs, various basis-penalty smooths could be used. In the applications of Section \ref{sec:examples}, we considered thin plate regression splines, which are optimal in the sense defined by \cite{wood2003}, with a shrinkage penalty to ensure that the smooth terms shrink to zero when the penalty tends to infinity \citep{marra2011}.

\subsubsection{Implementation}
\label{sec:model_implementation}

The marginal likelihood can be implemented in the R package TMB, which uses the Laplace approximation to integrate over the random effects \citep{kristensen2016}, and the design matrices for the basis functions and the penalty matrices can be computed with the R package mgcv \citep{wood2017}. A numerical optimiser (e.g., \texttt{optim} or \texttt{nlminb}) can then be used to minimise the marginal likelihood, and obtain estimates of the smoothness parameter $\bm\lambda$ and fixed effects $\bm\alpha$. We can get predicted values for the random effects $\bm\beta$, analogous to best linear unbiased predictors in linear mixed effect models, to infer the smooth relationships between SDE parameters and covariates. The joint precision matrix of fixed and random effects can be used for uncertainty quantification.

TMB makes it relatively simple to include other random effects in the model. It is often the case, e.g.\ in animal movement or financial studies, that the data arise from multiple instances of the SDE (multiple animals, stocks, etc.) and one wishes to fit a model that combines these instances while also allowing for inter-individual variation. The case of i.i.d.\ normal random effects is easily handled, as it is another type of basis-penalty smoother, where the penalty matrix is the identity matrix \citep[][Section 7.7]{wood2017}.

We describe the details of the implementation of this method, using mgcv and TMB, in Appendix B. We ran simulation experiments to investigate the performance of the proposed approach to recover the relationship between the SDE parameters and the covariates, under several model formulations. In those simulations, we thinned the simulated data to irregular time intervals, to mimic a real data set, and the method performed well in all scenarios (Appendix C). We also performed a simulation experiment to check the coverage of confidence intervals derived for $\bm\theta_t$ using the precision matrix given by TMB, and found that they correctly represented the uncertainty in the estimates (Appendix C).

\subsection{Model selection and model checking}

In this framework, it might be useful to discriminate between competing model formulations, e.g., different forms of the drift and diffusion terms. The problem of model selection in models involving basis-penalty smooths is relatively understudied outside standard GAMs. \cite{wood2017} describes two versions of the Akaike Information Criterion (AIC), the \emph{marginal} AIC and the \emph{conditional} AIC, based on different forms of the likelihood and AIC penalty (i.e., number of parameters in basic AIC). The marginal AIC uses the marginal likelihood defined in Eq.\ \ref{eq:marginal} with a penalty for the number of fixed effects $\bm\alpha$ and smoothing parameters $\bm\lambda$, and would be straightforward to implement in the framework of Section \ref{sec:model_implementation}. The conditional AIC is based on the joint penalised likelihood of fixed and random effects (Eq.\ \ref{eq:penlk}), with an additional AIC penalty on the complexity of smooth terms \citep[given by the number of effective degrees of freedom; see Section 5.4.2 of][]{wood2017}. For more detail about the respective limitations of these two criteria and possible solutions, see Section 6.11 of \cite{wood2017}. An alternative approach for model selection is to include an additional penalty in the likelihood, so that model components can be shrunk to zero (i.e., removed from the model) as part of the smoothness parameter estimation \citep{marra2011}.

For a chosen formulation, we propose a simple diagnostic to investigate goodness-of-fit in varying-coefficient SDE models. Based on the Euler-Maruyama discretisation of the process, a natural choice for model residuals $\epsilon_i$ is
\begin{linenomath*}
\begin{equation*}
  \epsilon_i = \dfrac{z_{i+1} - (z_i + \mu(z_i, \hat{\bm\theta}_{t_i}) \Delta_i)}{\sigma(z_i, \hat{\bm\theta}_{t_i}) \sqrt{\Delta_i}},
\end{equation*}
\end{linenomath*}
for $i = 1, \dots, n-1$, using the notation of Section \ref{sec:model_lk}, and where $\hat{\bm\theta}_{t_i}$ is the estimate of $\bm\theta_t$ over $[t_i, t_{i+1})$. Under the assumptions of the model (and of the discretisation), the residuals should be independent and approximately follow a standard normal distribution. In the analysis of Section \ref{sec:examples3}, we use quantile-quantile plots of the residuals to investigate lack of fit, and autocorrelation function plots to identify residual autocorrelation.

\section{Illustrative examples}
\label{sec:examples}

In this section, we present three analyses based on ecological data, to illustrate different applications of the models presented in Section \ref{sec:model}. We stress, however, that the varying-coefficient approach is general to SDE modelling, and our focus is chosen only because we are most familiar with these ecological problems. To further demonstrate the generality of the method, we also provide the analysis of a financial data set of oil prices in Appendix D.

\subsection{Linking elephant movement to environmental conditions}
\label{sec:examples1}

We illustrate the utility of varying-coefficient SDEs to analyse animal movement data, using the trajectory of an African elephant (\emph{Loxodonta africana}) presented by \cite{wall2014} and available on the Movebank data repository \citep{wall2014movebank}. We restricted the analysis to a period from May to September 2009 to avoid seasonal effects. The data set consisted of a time series of 3652 Easting-Northing locations, and also included the air temperature measured by the tag, at a time resolution of 1 hour (with a few missing observations).

We used a varying-coefficient version of the continuous-time correlated random walk model presented by \cite{johnson2008}; the original model has been used extensively to analyse animal location data. In this model, the (unobserved) velocity $\bm{V}_t$ of the animal is formulated as a varying-coefficient Ornstein-Uhlenbeck process,
\begin{linenomath*}
\begin{equation*}
    d\bm{V}_t = -r_t \bm{V}_t \mathop{dt} + s_t \mathop{d\bm{W}_t},
\end{equation*}
\end{linenomath*}
where $r_t$ and $s_t$ can be linked to the speed and sinuosity of the movement. This is a special case of the varying-coefficient SDE of Eq.\ \ref{eq:SDE2} where $\mu(\bm{V}_t, t) = -r_t \bm{V}_t$ and $\sigma(\bm{V}_t, t) = s_t$, i.e., with parameters $\bm\theta_t = (r_t, s_t)$. (Although the process $\bm{V}_t$ is bivariate, it is isotropic and the two dimensions can therefore be treated as two univariate processes driven by the same parameters.) Because the velocity is unobserved, the model can be written as a state-space model where $\bm{V}_t$ is latent, and where the observed process is the location of the animal, obtained as $\bm{Z}_t = \bm{Z}_0 + \int_0^t \bm{V}_s \mathop{ds}$. As described in Section \ref{sec:model_lk}, we implemented the likelihood using a Kalman filter with time-varying parameters. To investigate the effects of environmental conditions on the elephant's behaviour, we estimated the parameters $r_t$ and $s_t$ of the velocity process as functions of the air temperature. For interpretation, we then derived the parameter $\nu_t = \sqrt{\pi} s_t / (2 \sqrt{r_t})$, described by \cite{gurarie2017} as a measure of the speed of movement of the animal. Model fitting took about 5 min on a 1.3GHz Intel i7 CPU.

\begin{figure}[htbp]
    \centering
    \includegraphics[width = 0.8\textwidth]{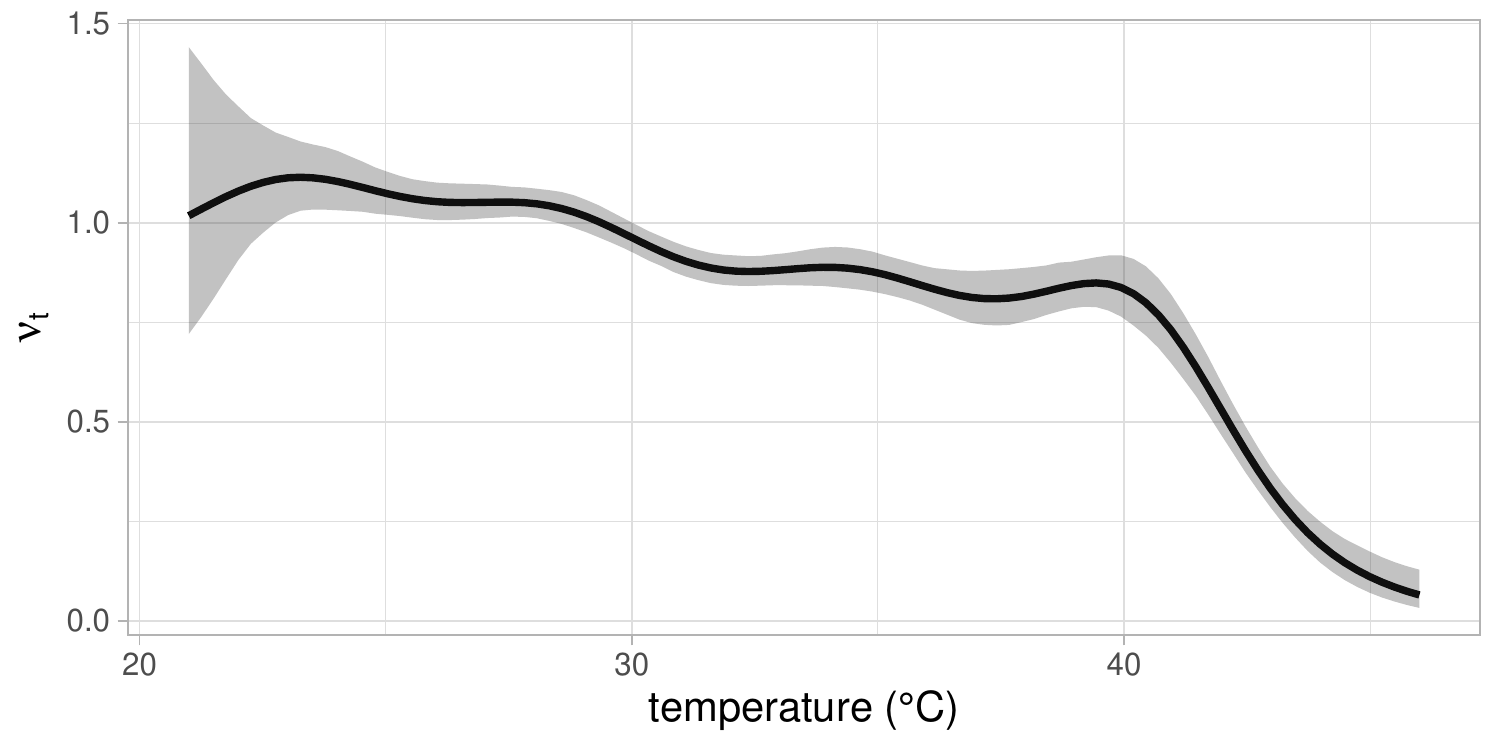}
    \caption{Results of the elephant analysis. Estimates of the speed parameter $\nu_t = \sqrt{\pi} s_t / (2 \sqrt{r_t})$, as a function of temperature. The black line is the mean estimate, and the grey shaded area is a 95\% confidence band.}
    \label{fig:eleph_res}
\end{figure}

Figure \ref{fig:eleph_res} shows estimates of the speed parameter $\nu_t$ as a function of temperature. A small value of $\nu_t$ corresponds to slow movement, encompassing behaviours with little activity (e.g., resting), and a large value corresponds to more active behaviours (e.g., exploration, transit). The speed parameter was highest at low temperatures (20-30 degrees), and decreased for higher temperatures, with a very steep decline above 40 degrees. This is consistent with what is known of the species: elephants are very sensitive to heat, and spend much of their time resting or waiting in the shade during periods of high temperatures \citep{mole2016}. Our non-parametric approach illuminated the non-linear relationship between the temperature and the speed of movement of this elephant. 

Other varying-coefficient models have been proposed in movement ecology to capture the effects of time-varying covariates on animal behaviour. In particular, \cite{hanks2015} modelled animal movement over a discrete spatial grid as a continuous-time Markov chain with time-varying transition rates, and linked transition rates to covariates using basis functions (similarly to Eq.\ \ref{eq:spline}). This continuous-time discrete-space model and its extensions have, e.g., been used to investigate the effects of time-varying environmental conditions on the movements of cougars \citep{hanks2015, buderman2018}, fur seals, and ants \citep{hanks2016}. The varying-coefficient SDEs presented in this paper offer an alternative framework to incorporate similar covariate effects in the dynamics of continuous-valued random processes (e.g., the continuous location or velocity of an animal).

SDEs have also been popular for analysing animal tracking data, e.g., to model animal behaviour \citep{blackwell1997}, the effects of environmental features on movement decisions \citep{preisler2004, michelot2019langevin}, and the emergence of home ranges \citep{dunn1977}. Regime-switching SDEs have been developed to allow for time-varying dynamics in the movement, where the latent state represents the behaviour of the animal \citep{blackwell1997, michelot2019ctcrw}. However, discrete behavioural states may lack the flexibility to capture the wide range of behaviours that animals display. It has also been difficult to include general covariate effects in that context; inference has typically required computationally-costly custom algorithms \citep{blackwell2016}. The method we propose to include factor covariates, linear or smooth effects of continuous covariates, and random effects in SDE models is an important step forward to link animal movement behaviour to environmental and individual-specific conditions. 

The output of regime-switching models (classification of data into clusters) may sometimes be more readily interpretable than the smoothly-varying parameters suggested here. In such cases, we could use a clustering algorithm on the estimated $\bm\theta_t$ values to identify different regimes in the time series, and interpret each cluster based on its centre, say. This procedure could be repeated on posterior draws of $\bm\theta_t$ to account for uncertainty in the clustering.

\cite{preisler2004} and \cite{russell2018} presented application-specific methods to define smooth relationships between movement parameters and spatial covariates in SDEs. The approach presented in this paper is a generalisation of their work to a wider class of SDEs, where any parameters can be specified using basis-penalty smooths. Note that the varying-coefficient CTCRW model could be fitted with the R package crawl \citep{johnson2008, johnson2018crawl}, using the joint likelihood of all parameters rather than the marginal likelihood of Eq.\ \ref{eq:marginal}. That package does not implement the smoothness parameter estimation and this would need to be done in an additional model selection stage.

\subsection{Body condition of elephant seals}
\label{sec:examples2}

We considered a study of body condition of elephant seals described by \cite{schick2013}, where the authors modelled body fat content over time, and how it was affected by environmental conditions. We used the data set from \cite{pirotta2019}, which includes information about drift dives of 26 Northern elephant seals  (\emph{Mirounga angustirostris}). The goal of the study was to investigate the dynamics of the body fat content of seals during migratory foraging trips, which last several months. Animals' body fat content cannot be observed directly when they are at sea. However, using telemetry tags fitted with depth sensors, we can measure the rate at which seals drift vertically in the water column during non-active dives \citep[``drift dives'';][]{biuw2003}. This drift rate is linked to the percentage of body fat because fat content affects buoyancy. A natural modelling approach, proposed by  \cite{schick2013}, is therefore to treat body fat content as a latent process in a state-space model, and estimate how it changes during a foraging trip from the drift rate observations.

We formulated a continuous-time analogue of the model of \cite{schick2013}, and defined the body fat content $L_t$ as a Brownian motion with time-varying drift, i.e., $dL_t = r_t \mathop{dt} \mathop{+} \sigma \mathop{dW_t}$ (Eq.\ \ref{eq:SDE2} with $\mu(L_t, t) = r_t$ and $\sigma(L_t, t) = \sigma$). The time-varying parameter $r_t$ measured the daily rate of change of the lipid content, with larger values indicating faster accumulation of fat mass. We combined the above SDE with the observation equation proposed by \cite{schick2013} to obtain the following state-space model formulation
\begin{linenomath*}
\begin{align*}
    \text{Observation process}\quad & D_i \sim 
        N \left( \alpha_1 + \alpha_2 \dfrac{L_i}{R_i},\ \dfrac{\tau^2}{h_i} \right) \\
    \text{State process}\quad & L_{i+1} \sim N(L_i + r_i \Delta_i,\ \sigma^2 \Delta_i)
\end{align*}
\end{linenomath*}
where $i$ is the day index, $D_i$ is the mean drift rate, $L_i$ is the lipid content, $R_i$ is the non-lipid content, $h_i$ is the number of drift dives, and $\alpha_1$, $\alpha_2$ and $\tau$ are parameters of the observation process. We followed \cite{schick2013} in assuming a constant diffusion $\sigma$, and investigated the effects of two covariates on the lipid change rate $r_t$: (1) surface transit per day, and (2) distance to the colony where the animals were tagged. We chose these covariates to link fat gains (i.e., foraging behaviour) to movement patterns and geographical location. In preliminary analyses, we included two other covariates from \cite{schick2013} (body fat proportion at departure, and daily number of drift dives), but found no evidence of an effect. We included a random normal intercept in $r_t$ to account for differences between seals.

We implemented the likelihood of this model with the Kalman filter, and estimated the effects of the covariates on the drift parameter following Section \ref{sec:model}. \cite{schick2013} fitted their state-space model within a Bayesian framework, using informative priors for $\tau^2$ and $\sigma^2$ based on biological knowledge. We included the same prior distributions as multiplicative terms in the likelihood, therefore performing maximum posterior estimation for those parameters. Model fitting took about 2 min on a 1.3GHz Intel i7 CPU. 

\begin{figure}[htbp]
    \centering
    \includegraphics[width=0.8\textwidth]{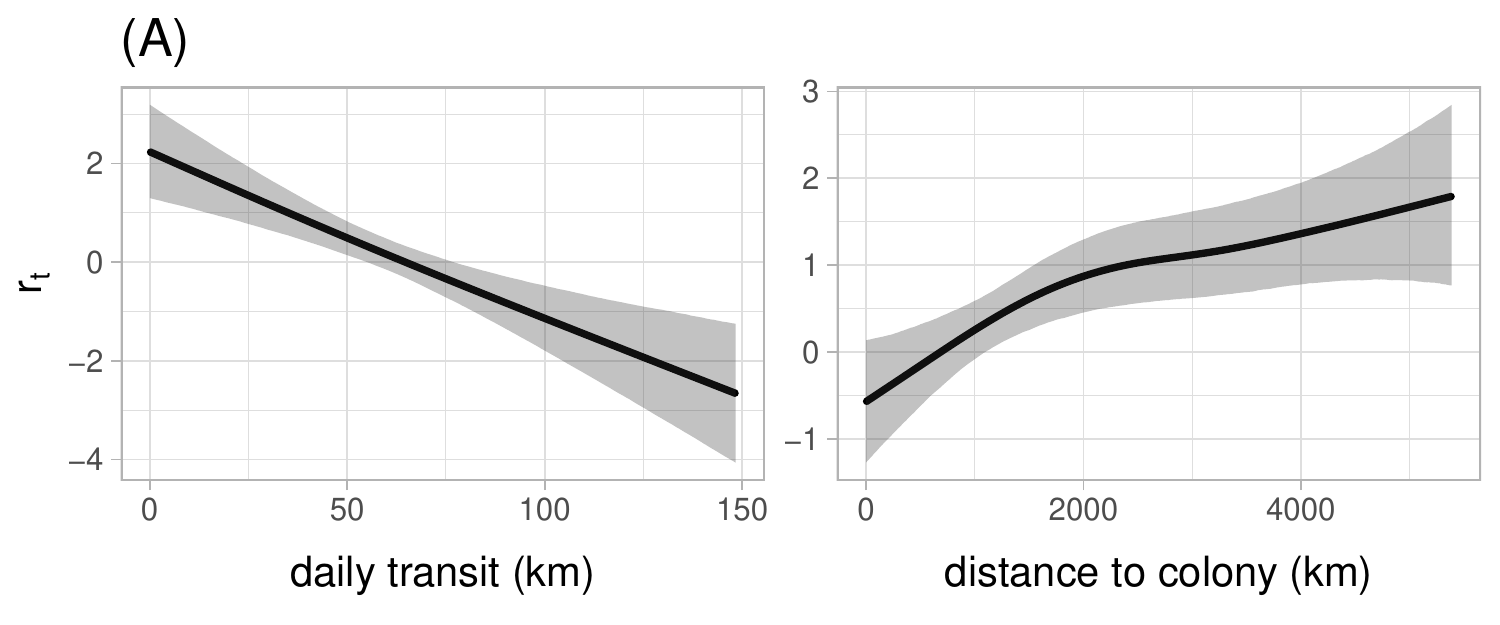}
    \includegraphics[width=0.8\textwidth]{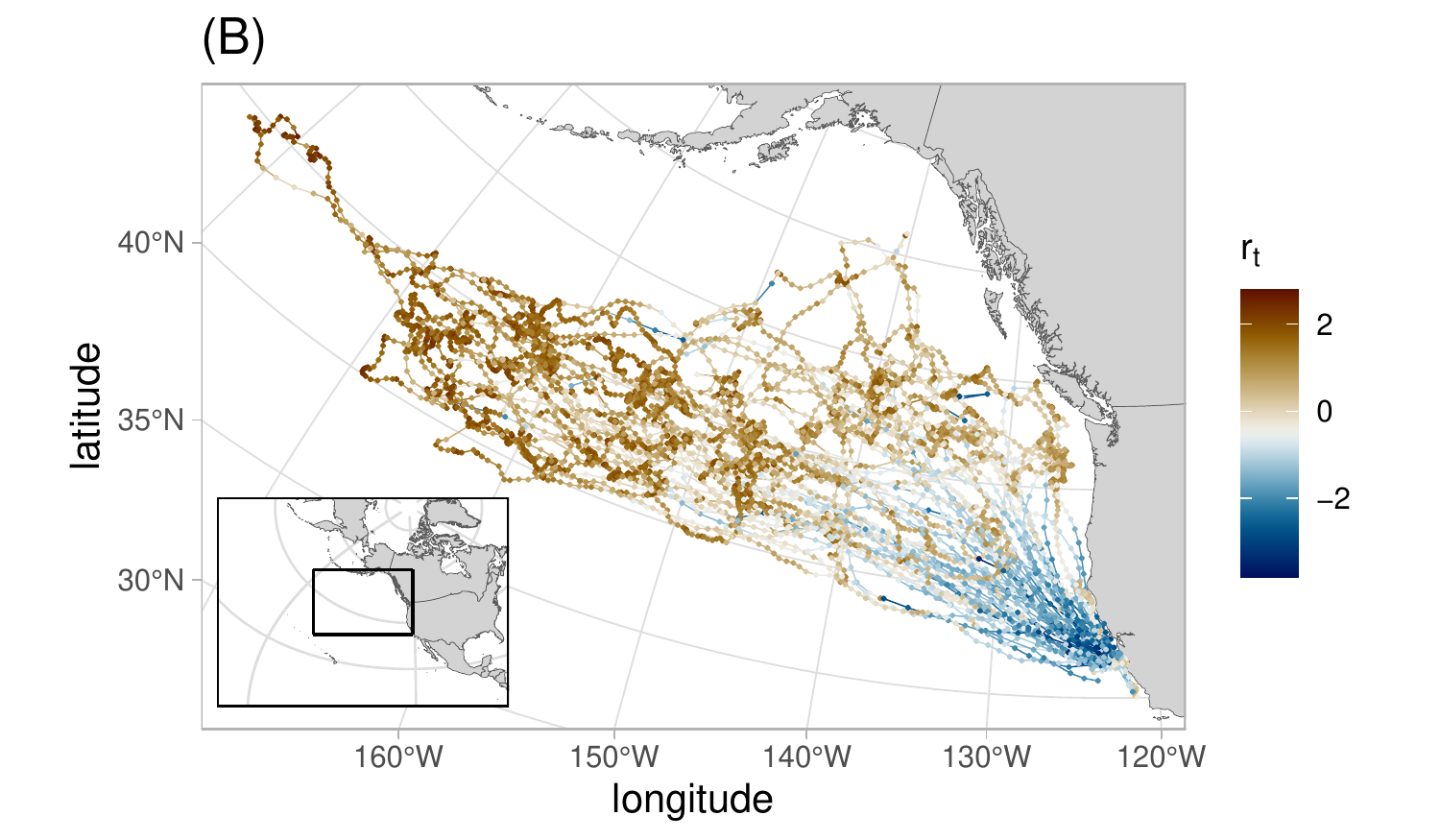}
    \caption{Results of elephant seal body condition study. (A) Estimated relationship between lipid gain parameter $r_t$ and two covariates: daily transit distance (left), and distance to colony (right). The black lines are mean estimates, and the grey shaded areas are 95\% confidence bands. Each estimate was obtained by fixing the other covariate to its mean. (B) Elephant seal movement tracks, coloured by predicted value of $r_t$. This figure appears in color in the electronic version of this article.}
    \label{fig:e_seals_res}
\end{figure}

Results are shown in Figure \ref{fig:e_seals_res}. The lipid gain rate $r_t$ was estimated to decrease with daily transit distance, which is consistent with the findings of \cite{schick2013}. This suggests that lipid gains are low when seals are travelling at high speeds, and that foraging is characterized by less horizontal movement. We also found that lipid gains increased with distance to the colony, in particular between 0 and 2000km. This indicates that animals must travel a considerable distance from their breeding colony to find foraging grounds that are rich enough for them to start gaining fat. Figure \ref{fig:e_seals_res} shows a map of the movement tracks of the seals, coloured by the predicted value of $r_t$, which highlights portions of the trips with high lipid gains. Our results provide a mechanistic justification for the assumption often made in elephant seal studies that slow horizontal movement at sea is associated with foraging behaviour \citep[e.g.,][]{michelot2017}.

This application shows how SDEs can be built as alternatives to discrete-time models \citep[such as the state-space model of][]{schick2013}, which do not depend on the time resolution of the data and can be applied to data collected at irregular intervals. Another difference with \cite{schick2013} is that we implemented the Kalman algorithm and Laplace approximation (with TMB) to integrate over latent components of the model, whereas they used computationally-costly Markov chain Monte Carlo methods.

\subsection{Diving behaviour of beaked whales}
\label{sec:examples3}

Beaked whales are marine mammals that routinely dive to depths in excess of 1km for periods of over an hour. Animal-borne telemetry tags allow study of their diving behaviour \citep{johnson2003, deruiter2013}. Here we consider data collected from high-resolution tags that include accelerometer and magnetometer sensors \citep[``DTAGs'';][]{johnson2003}, attached to four Cuvier's beaked whales (\emph{Ziphius cavirostris}). In general, beaked whales display two different types of dives with different physiological functions: deep and shallow dives. The structures of deep and shallow dives are very distinct, and would require separate models. Here, we focused on shallow dives, and also excluded sections of the data where the animals were at the sea surface (depth $<$ 15 m). The data set comprised $n_d = 73$ shallow dives from the four whales, with a median duration of 23 min. Multiple variables can be derived from DTAG data, and we computed the Euler angles (pitch, roll, and heading), which describe the posture of the animal in the water \citep{johnson2003}. The pitch is the angle between the main body axis and the horizontal, the roll is the angle around the main body axis, and the bearing is the angle in the horizontal plane (Figure S4 of supplementary material). The sampling rate of the raw data varied between 5Hz and 25Hz, and we downsampled by taking averages over non-overlapping 5-sec windows, to reduce the computational cost while keeping a sufficiently fine resolution to detect behavioural changes over each dive. This resulted in a total of 20041 observations for each variable.

Visual inspection of the data suggested that shallow dives all had a similar structure, with different phases of each dive displaying different levels of activity. Our aim was therefore to characterise the typical behaviour of beaked whales, as measured by their postural dynamics, during the different diving phases (e.g., descent, ascent, bottom). In preliminary analyses, we tried modelling each variable (pitch, roll, heading) with Brownian motion, but residual analysis revealed that the model did not capture heavy tails in increments of the process. We therefore replaced the Gaussian transition density with a generalized t distribution with fixed degrees of freedom ($\nu = 3$, based on visual data exploration), and estimated the location parameter $r_t$ and the scale parameter $s_t$ as time-varying. Here, we used the Euler-Maruyama discretization of the process (rather than the SDE itself) as the ``model of record'' to build a more complex model, as suggested by \cite{brillinger2010} in a similar context. Although this model is not a special case of the SDE given in Eq.\ \ref{eq:SDE2}, the method described in Section \ref{sec:model} can be applied, with the likelihood defined the pdf of a t distribution for each observed transition. The details of the model are described in Appendix E of the supplementary material. For this example, the three processes were treated as independent. 

\begin{figure}[htbp]
  \centering
  \includegraphics[width=\textwidth]{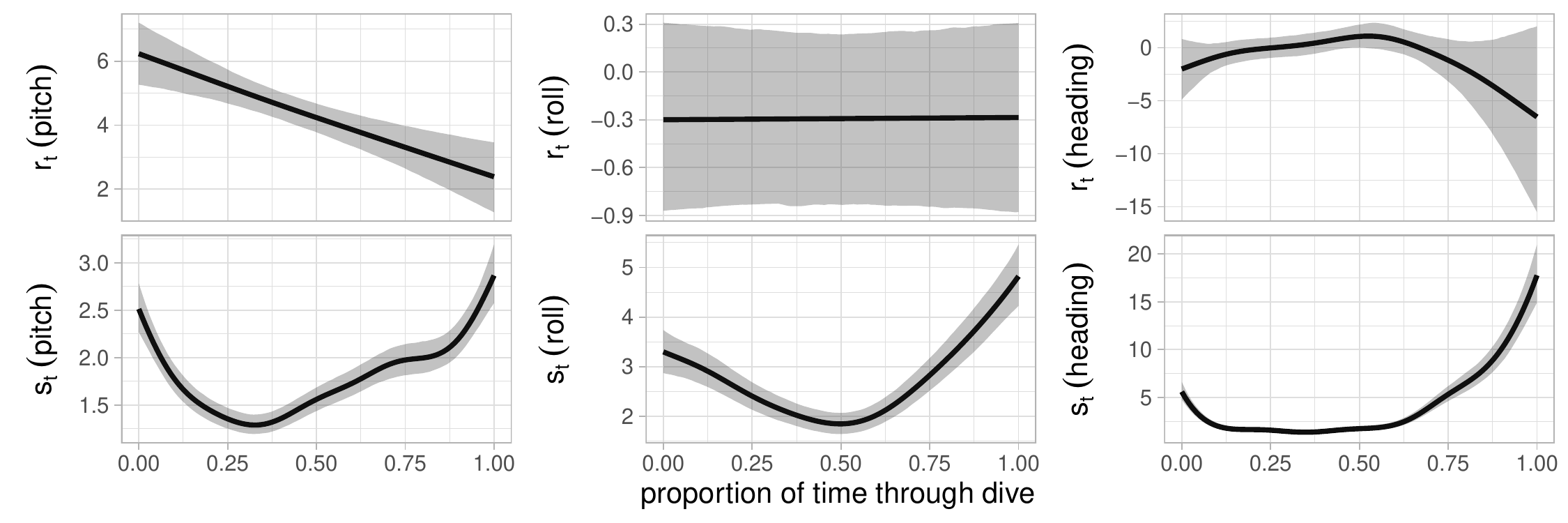}
  \caption{Estimates of the drift ($r_t$, top row) and diffusion ($s_t$, bottom row) parameters as functions of the proportion of time through the dive, in the beaked whale analysis. The three columns correspond to the three modelled variables: pitch (left), roll (middle), and heading (right). The black lines are the mean estimates, and the grey shaded areas are 95\% confidence bands. All estimates were obtained with the mean random effect intercept.}
  \label{fig:zc_est}
\end{figure}

The model had two parameters: the location $r_t$ and scale $s_t$ of the t-distributed increments. To investigate the time-varying behaviour of beaked whales, we specified $r_t$ and $s_t$ as smooth functions of the proportion of time through the dive $x_t \in [0, 1]$. We treated the dives as independent, and included random intercepts for the dive in $r_t$ and $s_t$, to account for variability between individuals and between dives. In summary, there were six time-varying parameters ($r_t$ and $s_t$ for each of the three data variables), modelled for each variable as
\begin{linenomath*}
\begin{align*}
    r_t & = \zeta_{d_t} + f_r(x_t),& \zeta_j \sim N(\mu_\zeta, (\sigma_\zeta)^2) \text{ for } j \in \{ 1, 2, \dots, n_d \}, \\
    \log(s_t) & = \xi_{d_t} + f_s(x_t),& \xi_j \sim N(\mu_\xi, (\sigma_\xi)^2) \text{ for } j \in \{ 1, 2, \dots, n_d \},
\end{align*}
\end{linenomath*}
where $d_t \in \{ 1, 2, \dots, n_d \}$ is the dive index at time $t$, $f_r$ and $f_s$ are basis-penalty smooths, and $\{\mu_\zeta, \sigma_\zeta, \mu_\xi, \sigma_\xi \}$ are unknown hyper-parameters. Model fitting took 30 min on a 1.3GHz Intel i7 CPU. The estimated relationships between the parameters and the proportion of time through the dive are shown in Figure \ref{fig:zc_est}.

The estimated drift parameter $r_t$ for pitch was positive over the whole dive, suggesting that pitch tended to increase during a typical dive. This is consistent with the observed convex shape of the dives: pitch increases between the descent and bottom phases, and again between the bottom and ascent phases. The estimated drift for roll was close to zero, and did not seem to be affected by the phase of the dive, suggesting that there were no particular trend in that variable. The estimated drift in heading was negative during the final part of the dive (ascent), but the confidence bands were wide and overlapped zero. All three diffusion parameters $s_t$ suggested that there was more variability at the start and end of each dive, i.e. during ascent and descent, than when the whale was at the bottom. This variability can be viewed as a proxy for the level of activity: more diffusion suggests more frequent postural changes. Variability in pitch was low during the bottom phase, which may be associated with gliding motion, whereas it was high during descent and ascent, suggesting continued stroking or ``stroke-and-glide'' motion. These changes in the pitch diffusion parameter showed how whales alternate between different swimming styles over each dive, which has been linked to energetic efficiency in response to drag forces and buoyancy \citep{miller2004, lopez2015}. Roll displayed highest variability during the ascent phase, and the diffusion parameter for heading was much higher during the final phase of the dive, just before the whales surfaced again, corresponding to more directional changes in the horizontal plane. These postural changes may have several functions, such as locating predators before surfacing (when the whales are most vulnerable), socialising with conspecifics, and orienting to sea currents.

\begin{figure}[htbp]
    \centering
    \includegraphics[width=0.32\textwidth]{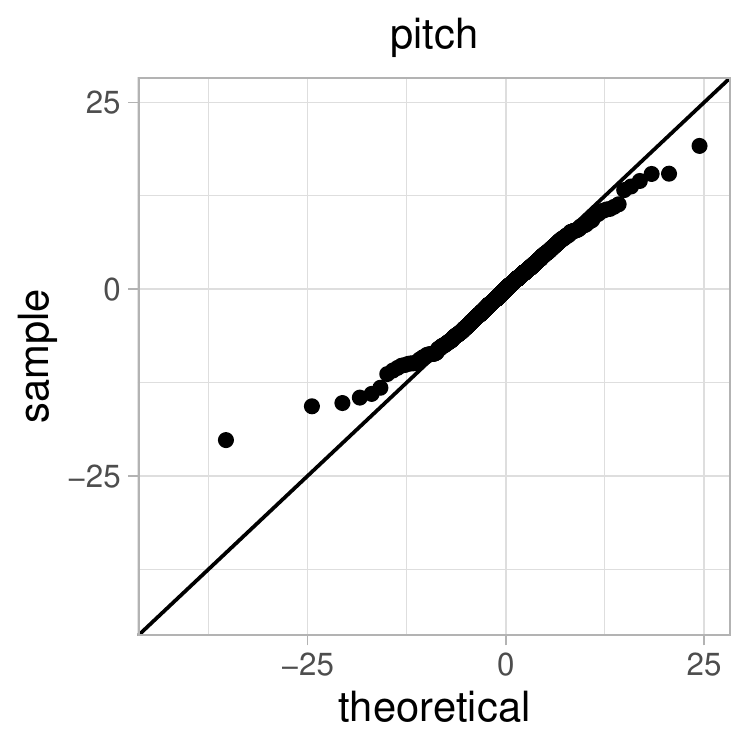}
    \includegraphics[width=0.32\textwidth]{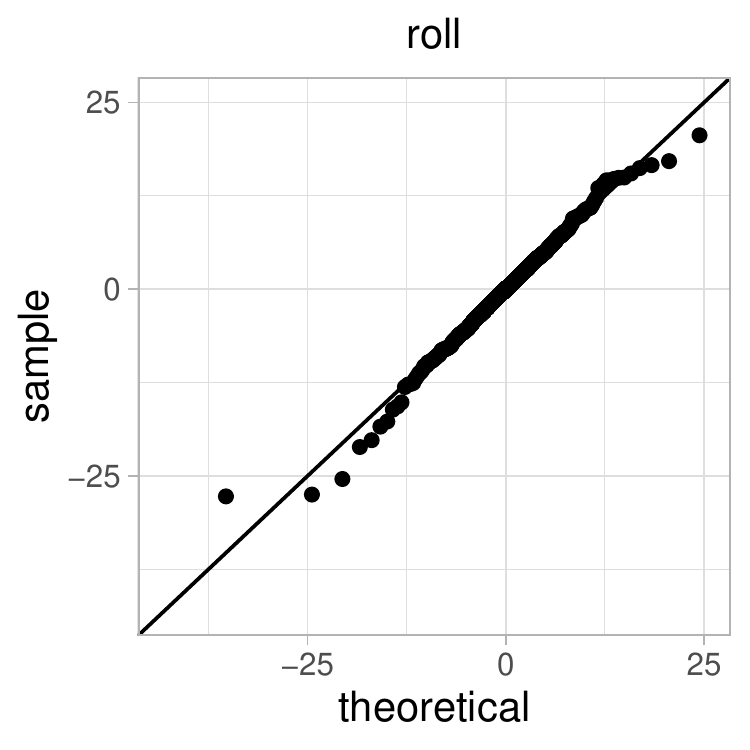}
    \includegraphics[width=0.32\textwidth]{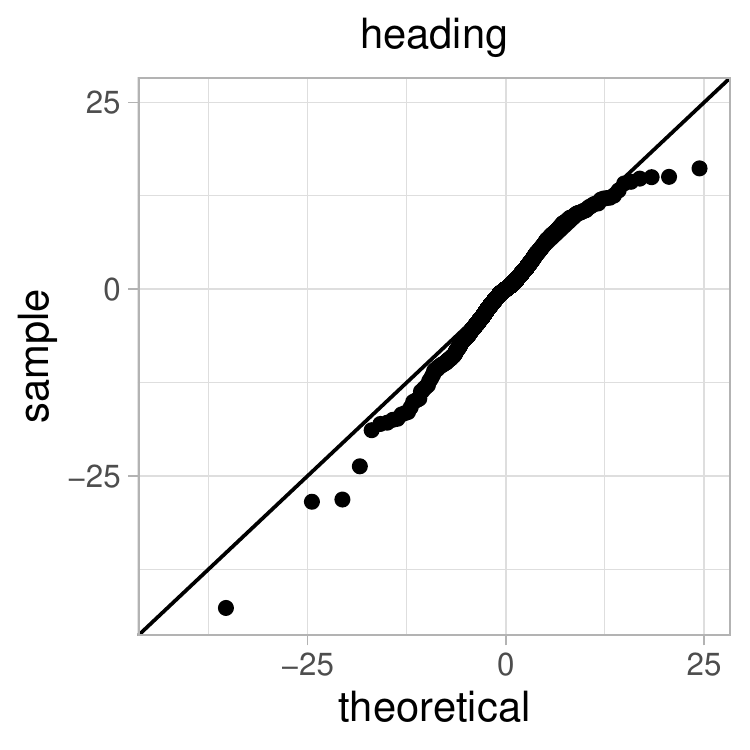}
    \caption{Quantile-quantile plots of the residuals against the Student's t distribution with $\nu = 3$ degrees of freedom, for the beaked whale analysis.}
    \label{fig:zc_res}
\end{figure}

Quantile-quantile plots of the residuals against the appropriate standardized t distributions are shown in Figure \ref{fig:zc_res}, and suggest appropriate fit. However, we found autocorrelation in the residuals using autocorrelation function plots, pointing to features of the whales' movement that were not included in the model (Figure S5 of the supplementary material). Most notably, there was positive autocorrelation in the heading residuals over about 1-2 minutes, suggesting that some persistence in heading was not captured by the model. It might be more adequate to model heading with a process that induces correlation between increments, such as a one-dimensional version of the continuous-time correlated random walk used in Section \ref{sec:examples1}. There was no correlation between residuals of the three processes (pitch, roll, and heading), supporting our decision to model them separately.

Most existing analyses of DTAG data have been based on dive-by-dive summary statistics of activity (e.g., dive duration, maximum depth), and have looked at broad behavioural patterns \citep[e.g.,][]{deruiter2013, quick2017}. This contrasts with our approach, where the behaviour of beaked whales is modelled at a fine time resolution over each dive, and the dives are treated as realisations of an underlying random process. Using varying-coefficient SDEs, we could estimate a more detailed description of the within-dive activity of whales. As another alternative, discrete-time regime-switching models have been proposed to analyse data of this kind \citep[i.e., hidden Markov models;][]{isojunno2015, leos2017}, and a similar continuous-time approach could be implemented. In that setting, the states of the latent process would represent discrete regimes of activity. For high-resolution data, however, it may often be preferable to model behaviour as changing smoothly in time (rather than switching between discrete states). The method that we use also makes it straightforward to investigate the effects of covariates (if available), and to include random effects to capture differences between individuals.

\section{Discussion}
\label{sec:discussion}

In this paper, we have focused on a univariate diffusion process $(Z_t)$, and suggested using independent diffusion processes for each dimension in the case of multivariate data (e.g., the three postural angles in Section \ref{sec:examples3}). The proposed method could similarly be applied to the $N$-dimensional diffusion process $(\bm{Z}_t)$ defined by the equation $d\bm{Z}_t = \bm\mu(\bm{Z}_t, \bm\theta_t) dt + \bm\sigma(\bm{Z}_t, \bm\theta_t) d\bm{W}_t$, where $\bm\mu(\bm{Z}_t, \bm\theta_t) \in \mathbb{R}^N$,  $\bm{W}_t \in \mathbb{R}^N$, and $\bm\sigma(\bm{Z}_t, \bm\theta_t) \in \mathbb{R}^{N \times N}$. In this case, the drift $\bm\mu$ and diffusion $\bm\sigma$ might be functions of several time-varying parameters. We can use the Euler-Maruyama discretization to obtain the (approximate) transition density as a multivariate normal distribution, and estimate the model parameters as in Section \ref{sec:model}. A simulation study may be required to investigate identifiability in such models, when a large number of time-varying parameters must be estimated jointly.

The method of inference that we presented in Section \ref{sec:model} is approximate, as it relies on the time discretization of the drift and diffusion functions. As suggested in Section \ref{sec:model_lk}, data augmentation can be used to improve the accuracy of the method, but there are no general guidelines as to when this might be necessary. Recent studies have evaluated the approximation error of the Euler-Maruyama method for special cases of SDEs \citep{albertsen2019, michelot2019langevin}, but the performance of the approach will be application-specific. The discretization is based on the assumption that the SDE parameters $\bm\theta_t$ are fixed over each time interval, i.e., the error depends on how fast $\bm\theta_t$ varies over time. Future work could focus on developing diagnostics to assess the time resolution of discretization, e.g.\ based on the distribution of first-order differences in estimated values of $\bm\theta_t$.

The parameters of a varying-coefficient SDE are specified using basis-penalty smooths, and we therefore assume a smooth relationship between parameters and covariates. There is some flexibility in this formulation, because the smoothness parameter is estimated from the data, but it may not always be appropriate. In particular, if the relationship involves abrupt changes (e.g., discontinuities), then it may not be well captured by a smooth function, and regime-switching models may be preferable. To better understand this issue, it would be interesting to compare the results of the varying-coefficient approach and a regime-switching model on the same data set. However, this may not be straightforward in practice, as there is no generally-applicable method to include covariates in continuous-time regime-switching models.
We could also consider integrating regime switches into varying-coefficient SDEs, i.e., defining each SDE parameter by several smooths between which the process switches through time. This model can be viewed as a continuous-time hidden Markov model where the state-dependent observation distribution is given by the transition density of the SDE, and the approximate likelihood of this model could be obtained using existing methodology.

We suggested using the package TMB to implement the marginal likelihood and to integrate over random effects using the Laplace approximation. TMB also uses automatic differentiation to evaluate the gradient of the log-likelihood, which improves computational speed \citep{kristensen2016}. This fast implementation has a downside: to build the gradient function, TMB needs to create the ``computational graph'' of the likelihood, i.e., its representation in terms of elementary functions (for which the analytical gradient is known). In our experience, the construction of this graph can be memory-intensive for large data sets or complex model formulations, and may not be feasible on standard desktop computers. In those cases, high-performance computing systems with more memory may be required.


The model presented for the time-varying parameters of SDEs relies on the general methodology of generalized additive models (GAMs), which has been greatly extended beyond the basic formulation presented herein. In particular, an interesting direction for future research will be the implementation of hierarchical GAMs \citep{pedersen2019} in this framework. Here, the smooth relationship between response and covariates can vary across groups, while retaining some common features (related to shape and degree of smoothness). This extension could be applied to investigate inter-individual differences in ecological analyses, with more nuance than the simple random-intercept model mentioned in Section \ref{sec:model_implementation}. We could, e.g., define the response of several animals to an environmental covariate with functions comprising a population-level mean component and individual-level components measuring the individual deviations from the mean. Other extensions of GAMs, such as adaptive smoothing \citep[][Section 5.3.5]{wood2017} or tensor product smooth interactions \citep[][Section 5.6]{wood2017}, could further increase the applicability of varying-coefficient SDEs.

\subsubsection*{Acknowledgments}

\emph{\footnotesize We would like to thank Simon Chamaill\'e-Jammes, Stacy DeRuiter, Alan Gelfand, Josh Hewitt, Dave Miller, Nicola Quick, and Rob Schick for fruitful discussions about this research, as well as the reviewers and editors for suggestions that greatly improved the paper. This work was funded by the US Office of Naval Research, grant N000141812807. The beaked whale data of Section \ref{sec:examples3} was collected as part of the SOCAL-BRS project, primarily funded by the US Navy’s Chief of Naval Operations Environmental Readiness Division and subsequently by the US Navy's Living Marine Resources Program. Additional support for environmental sampling and logistics was also provided by the Office of Naval Research, Marine Mammal Program. All research activities for that study were authorized and conducted under US National Marine Fisheries Service permit 14534; Channel Islands National Marine Sanctuary permit 2010-004; US Department of Defense Bureau of Medicine and Surgery authorization; a federal consistency determination by the California Coastal Commission; and numerous institutional animal care and use committee authorizations. We are grateful to Jake Wall, George Wittemyer, Valerie LeMay, Iain Douglas‐Hamilton and Brian Klinkenberg for making the elephant data used in Section \ref{sec:examples1} publicly available, and to Enrico Pirotta, Lisa Schwarz, Daniel Costa, Patrick Robinson and Leslie New for making the elephant seal data used in Section \ref{sec:examples2} publicly available.}

\bibliographystyle{apalike}
\bibliography{refs.bib}

\begin{thebibliography}{}

\bibitem[A{\"i}t-Sahalia and Kimmel, 2007]{ait2007}
A{\"i}t-Sahalia, Y. and Kimmel, R. (2007).
\newblock Maximum likelihood estimation of stochastic volatility models.
\newblock {\em Journal of Financial Economics}, 83(2):413--452.

\bibitem[Albertsen, 2019]{albertsen2019}
Albertsen, C.~M. (2019).
\newblock Generalizing the first-difference correlated random walk for marine
  animal movement data.
\newblock {\em Scientific Reports}, 9(1):1--14.

\bibitem[Albertsen et~al., 2015]{albertsen2015}
Albertsen, C.~M., Whoriskey, K., Yurkowski, D., Nielsen, A., and Flemming,
  J.~M. (2015).
\newblock Fast fitting of non-{Gaussian} state-space models to animal movement
  data via {Template} {Model} {Builder}.
\newblock {\em Ecology}, 96(10):2598--2604.

\bibitem[Allen, 2007]{allen2007}
Allen, E. (2007).
\newblock {\em Modeling with {I}t{\^o} stochastic differential equations},
  volume~22.
\newblock Springer Science \& Business Media.

\bibitem[Allen and Van~den Driessche, 2006]{allen2006}
Allen, L.~J. and Van~den Driessche, P. (2006).
\newblock Stochastic epidemic models with a backward bifurcation.
\newblock {\em Mathematical Biosciences \& Engineering}, 3(3):445.

\bibitem[Archambeau et~al., 2007]{archambeau2007}
Archambeau, C., Cornford, D., Opper, M., and Shawe-Taylor, J. (2007).
\newblock Gaussian process approximations of stochastic differential equations.
\newblock {\em Journal of Machine Learning Research}, 1:1--16.

\bibitem[Biuw et~al., 2003]{biuw2003}
Biuw, M., McConnell, B., Bradshaw, C.~J., Burton, H., and Fedak, M. (2003).
\newblock Blubber and buoyancy: monitoring the body condition of free-ranging
  seals using simple dive characteristics.
\newblock {\em Journal of Experimental Biology}, 206(19):3405--3423.

\bibitem[Blackwell, 1997]{blackwell1997}
Blackwell, P.~G. (1997).
\newblock Random diffusion models for animal movement.
\newblock {\em Ecological Modelling}, 100(1-3):87--102.

\bibitem[Blackwell et~al., 2016]{blackwell2016}
Blackwell, P.~G., Niu, M., Lambert, M.~S., and LaPoint, S.~D. (2016).
\newblock Exact {Bayesian} inference for animal movement in continuous time.
\newblock {\em Methods in Ecology and Evolution}, 7(2):184--195.

\bibitem[Brillinger, 2010]{brillinger2010}
Brillinger, D.~R. (2010).
\newblock Modeling spatial trajectories.
\newblock In Gelfand, A., Diggle, P., Guttorp, P., and Fuentes, M., editors,
  {\em Handbook of Spatial Statistics}, pages 463--474. CRC Press, Boca Raton,
  Florida, USA.

\bibitem[Buderman et~al., 2018]{buderman2018}
Buderman, F.~E., Hooten, M.~B., Alldredge, M.~W., Hanks, E.~M., and Ivan, J.~S.
  (2018).
\newblock Time-varying predatory behavior is primary predictor of fine-scale
  movement of wildland-urban cougars.
\newblock {\em Movement Ecology}, 6(1):22.

\bibitem[Dennis et~al., 1991]{dennis1991}
Dennis, B., Munholland, P.~L., and Scott, J.~M. (1991).
\newblock Estimation of growth and extinction parameters for endangered
  species.
\newblock {\em Ecological Monographs}, 61(2):115--143.

\bibitem[DeRuiter et~al., 2013]{deruiter2013}
DeRuiter, S.~L., Southall, B.~L., Calambokidis, J., Zimmer, W.~M., Sadykova,
  D., Falcone, E.~A., Friedlaender, A.~S., Joseph, J.~E., Moretti, D., Schorr,
  G.~S., et~al. (2013).
\newblock First direct measurements of behavioural responses by {Cuvier's}
  beaked whales to mid-frequency active sonar.
\newblock {\em Biology Letters}, 9(4):20130223.

\bibitem[Duan et~al., 2009]{duan2009}
Duan, J.~A., Gelfand, A.~E., Sirmans, C., et~al. (2009).
\newblock Modeling space-time data using stochastic differential equations.
\newblock {\em Bayesian Analysis}, 4(4):733--758.

\bibitem[Dunn and Gipson, 1977]{dunn1977}
Dunn, J.~E. and Gipson, P.~S. (1977).
\newblock Analysis of radio telemetry data in studies of home range.
\newblock {\em Biometrics}, 33(1):85--101.

\bibitem[Elerian et~al., 2001]{elerian2001}
Elerian, O., Chib, S., and Shephard, N. (2001).
\newblock Likelihood inference for discretely observed nonlinear diffusions.
\newblock {\em Econometrica}, 69(4):959--993.

\bibitem[Garc\'ia et~al., 2017]{garcia2017}
Garc\'ia, C.~A., Otero, A., Felix, P., Presedo, J., and Marquez, D.~G. (2017).
\newblock Nonparametric estimation of stochastic differential equations with
  sparse {Gaussian} processes.
\newblock {\em Physical Review E}, 96(2):022104.

\bibitem[Gurarie et~al., 2017]{gurarie2017}
Gurarie, E., Fleming, C.~H., Fagan, W.~F., Laidre, K.~L., Hern{\'a}ndez-Pliego,
  J., and Ovaskainen, O. (2017).
\newblock Correlated velocity models as a fundamental unit of animal movement:
  synthesis and applications.
\newblock {\em Movement Ecology}, 5(1):13.

\bibitem[Hanks et~al., 2015]{hanks2015}
Hanks, E.~M., Hooten, M.~B., Alldredge, M.~W., et~al. (2015).
\newblock Continuous-time discrete-space models for animal movement.
\newblock {\em The Annals of Applied Statistics}, 9(1):145--165.

\bibitem[Hanks and Hughes, 2016]{hanks2016}
Hanks, E.~M. and Hughes, D.~A. (2016).
\newblock Flexible discrete space models of animal movement.
\newblock {\em arXiv preprint arXiv:1606.07986}.

\bibitem[Hastie and Tibshirani, 1993]{hastie1993}
Hastie, T. and Tibshirani, R. (1993).
\newblock Varying-coefficient models.
\newblock {\em Journal of the Royal Statistical Society: Series B
  (Methodological)}, 55(4):757--779.

\bibitem[Isojunno and Miller, 2015]{isojunno2015}
Isojunno, S. and Miller, P.~J. (2015).
\newblock Sperm whale response to tag boat presence: biologically informed
  hidden state models quantify lost feeding opportunities.
\newblock {\em Ecosphere}, 6(1):1--46.

\bibitem[Johnson and London, 2018]{johnson2018crawl}
Johnson, D.~S. and London, J.~M. (2018).
\newblock crawl: an {R} package for fitting continuous-time correlated random
  walk models to animal movement data.

\bibitem[Johnson et~al., 2008]{johnson2008}
Johnson, D.~S., London, J.~M., Lea, M.-A., and Durban, J.~W. (2008).
\newblock Continuous-time correlated random walk model for animal telemetry
  data.
\newblock {\em Ecology}, 89(5):1208--1215.

\bibitem[Johnson and Tyack, 2003]{johnson2003}
Johnson, M.~P. and Tyack, P.~L. (2003).
\newblock A digital acoustic recording tag for measuring the response of wild
  marine mammals to sound.
\newblock {\em IEEE Journal of Oceanic Engineering}, 28(1):3--12.

\bibitem[Kristensen et~al., 2016]{kristensen2016}
Kristensen, K., Nielsen, A., Berg, C., Skaug, H., and Bell, B. (2016).
\newblock {TMB}: Automatic differentiation and {Laplace} approximation.
\newblock {\em Journal of Statistical Software}, 70(5):1--21.

\bibitem[Leos-Barajas et~al., 2017]{leos2017}
Leos-Barajas, V., Photopoulou, T., Langrock, R., Patterson, T.~A., Watanabe,
  Y.~Y., Murgatroyd, M., and Papastamatiou, Y.~P. (2017).
\newblock Analysis of animal accelerometer data using hidden {Markov} models.
\newblock {\em Methods in Ecology and Evolution}, 8(2):161--173.

\bibitem[Liechty and Roberts, 2001]{liechty2001}
Liechty, J.~C. and Roberts, G.~O. (2001).
\newblock Markov chain {Monte} {Carlo} methods for switching diffusion models.
\newblock {\em Biometrika}, 88(2):299--315.

\bibitem[Mao and Yuan, 2006]{mao2006}
Mao, X. and Yuan, C. (2006).
\newblock {\em Stochastic differential equations with Markovian switching}.
\newblock Imperial College Press.

\bibitem[Marra and Wood, 2011]{marra2011}
Marra, G. and Wood, S.~N. (2011).
\newblock Practical variable selection for generalized additive models.
\newblock {\em Computational Statistics \& Data Analysis}, 55(7):2372--2387.

\bibitem[Mart{\'\i}n~L{\'o}pez et~al., 2015]{lopez2015}
Mart{\'\i}n~L{\'o}pez, L.~M., Miller, P.~J., Aguilar~de Soto, N., and Johnson,
  M. (2015).
\newblock Gait switches in deep-diving beaked whales: biomechanical strategies
  for long-duration dives.
\newblock {\em Journal of Experimental Biology}, 218(9):1325--1338.

\bibitem[Michelot and Blackwell, 2019]{michelot2019ctcrw}
Michelot, T. and Blackwell, P.~G. (2019).
\newblock State-switching continuous-time correlated random walks.
\newblock {\em Methods in Ecology and Evolution}, 10(5):637--649.

\bibitem[Michelot et~al., 2019]{michelot2019langevin}
Michelot, T., Gloaguen, P., Blackwell, P.~G., and {\'E}tienne, M.-P. (2019).
\newblock The {Langevin} diffusion as a continuous-time model of animal
  movement and habitat selection.
\newblock {\em Methods in Ecology and Evolution}, 10(11):1894--1907.

\bibitem[Michelot et~al., 2017]{michelot2017}
Michelot, T., Langrock, R., Bestley, S., Jonsen, I.~D., Photopoulou, T., and
  Patterson, T.~A. (2017).
\newblock Estimation and simulation of foraging trips in land-based marine
  predators.
\newblock {\em Ecology}, 98(7):1932--1944.

\bibitem[Miller et~al., 2004]{miller2004}
Miller, P.~J., Johnson, M.~P., Tyack, P.~L., and Terray, E.~A. (2004).
\newblock Swimming gaits, passive drag and buoyancy of diving sperm whales
  {Physeter} macrocephalus.
\newblock {\em Journal of Experimental Biology}, 207(11):1953--1967.

\bibitem[Mole et~al., 2016]{mole2016}
Mole, M.~A., Rodrigues~D{\'A}raujo, S., Van~Aarde, R.~J., Mitchell, D., and
  Fuller, A. (2016).
\newblock Coping with heat: behavioural and physiological responses of savanna
  elephants in their natural habitat.
\newblock {\em Conservation Physiology}, 4(1).

\bibitem[Pedersen et~al., 2019]{pedersen2019}
Pedersen, E.~J., Miller, D.~L., Simpson, G.~L., and Ross, N. (2019).
\newblock Hierarchical generalized additive models in ecology: an introduction
  with mgcv.
\newblock {\em PeerJ}, 7:e6876.

\bibitem[Pirotta et~al., 2019]{pirotta2019}
Pirotta, E., Schwarz, L.~K., Costa, D.~P., Robinson, P.~W., and New, L. (2019).
\newblock Modeling the functional link between movement, feeding activity, and
  condition in a marine predator.
\newblock {\em Behavioral Ecology}, 30(2):434--445.

\bibitem[Preisler et~al., 2004]{preisler2004}
Preisler, H.~K., Ager, A.~A., Johnson, B.~K., and Kie, J.~G. (2004).
\newblock Modeling animal movements using stochastic differential equations.
\newblock {\em Environmetrics}, 15(7):643--657.

\bibitem[Quick et~al., 2017]{quick2017}
Quick, N.~J., Isojunno, S., Sadykova, D., Bowers, M., Nowacek, D.~P., and Read,
  A.~J. (2017).
\newblock Hidden {Markov} models reveal complexity in the diving behaviour of
  short-finned pilot whales.
\newblock {\em Scientific Reports}, 7(1):1--12.

\bibitem[Rajabzadeh et~al., 2016]{rajabzadeh2016}
Rajabzadeh, Y., Rezaie, A.~H., and Amindavar, H. (2016).
\newblock A robust nonparametric framework for reconstruction of stochastic
  differential equation models.
\newblock {\em Physica A: Statistical Mechanics and its Applications},
  450:294--304.

\bibitem[Russell et~al., 2018]{russell2018}
Russell, J.~C., Hanks, E.~M., Haran, M., Hughes, D., et~al. (2018).
\newblock A spatially varying stochastic differential equation model for animal
  movement.
\newblock {\em The Annals of Applied Statistics}, 12(2):1312--1331.

\bibitem[Schick et~al., 2013]{schick2013}
Schick, R.~S., New, L.~F., Thomas, L., Costa, D.~P., Hindell, M.~A., McMahon,
  C.~R., Robinson, P.~W., Simmons, S.~E., Thums, M., Harwood, J., et~al.
  (2013).
\newblock Estimating resource acquisition and at-sea body condition of a marine
  predator.
\newblock {\em Journal of Animal Ecology}, 82(6):1300--1315.

\bibitem[Uhlenbeck and Ornstein, 1930]{uhlenbeck1930}
Uhlenbeck, G.~E. and Ornstein, L.~S. (1930).
\newblock On the theory of the {Brownian} motion.
\newblock {\em Physical Review}, 36(5):823.

\bibitem[Vasicek, 1977]{vasicek1977}
Vasicek, O. (1977).
\newblock An equilibrium characterization of the term structure.
\newblock {\em Journal of Financial Economics}, 5(2):177--188.

\bibitem[Wahba, 1990]{wahba1990}
Wahba, G. (1990).
\newblock {\em Spline models for observational data}.
\newblock SIAM, Philadelphia.

\bibitem[Wall et~al., 2014a]{wall2014movebank}
Wall, J., Wittemyer, G., LeMay, V., Douglas-Hamilton, I., and Klinkenberg, B.
  (2014a).
\newblock Data from: {Elliptical} time-density model to estimate wildlife
  utilization distributions.
\newblock Movebank data repository, DOI:10.5441/001/1.f321pf80/1.

\bibitem[Wall et~al., 2014b]{wall2014}
Wall, J., Wittemyer, G., LeMay, V., Douglas-Hamilton, I., and Klinkenberg, B.
  (2014b).
\newblock Elliptical time-density model to estimate wildlife utilization
  distributions.
\newblock {\em Methods in Ecology and Evolution}, 5(8):780--790.

\bibitem[Wood, 2003]{wood2003}
Wood, S.~N. (2003).
\newblock Thin plate regression splines.
\newblock {\em Journal of the Royal Statistical Society: Series B (Statistical
  Methodology)}, 65(1):95--114.

\bibitem[Wood, 2017]{wood2017}
Wood, S.~N. (2017).
\newblock {\em Generalized additive models: an introduction with {R}}.
\newblock CRC press.
\newblock Second Edition.

\end{thebibliography}


\setcounter{equation}{0}
\renewcommand\thefigure{S\arabic{figure}}
\setcounter{figure}{0}
\appendix
\newpage

\section*{Appendix A\quad Special cases with applications in ecology and finance}

The specification of the functions $\mu$ and $\sigma$ in the SDE determines the type of diffusion process (Equation 2 of the paper). In practice, this choice depends on the characteristics of the observed process. In this section, we propose several choices of $\mu$ and $\sigma$, corresponding to varying-coefficient variants of standard diffusion processes, and possible applications to the analysis of animal movement and other ecological data. In all the examples that we consider here, the drift function $\mu$ and the diffusion function $\sigma$ each depend on only one time-varying parameter, i.e.\ $\mu(Z_t, \bm\theta_t) = \mu(Z_t, \theta_t^{(1)})$ and $\sigma(Z_t, \bm\theta_t) = \sigma(Z_t, \theta_t^{(2)})$, where $\bm\theta_t = (\theta_t^{(1)}, \theta_t^{(2)})$. We denote $r_t = \theta_t^{(1)}$ and $s_t = \theta_t^{(2)}$ for simplicity.

\paragraph{Brownian motion with drift} The simplest model is the Brownian motion (with drift), where $\mu(Z_t, \bm\theta_t) = r_t$ and $\sigma(Z_t, \bm\theta_t) = s_t$. Here, $r_t$ and $s_t$ are time-varying drift and diffusion parameters for the process, respectively. Based on the Euler-Maruyama discretization, the approximate transition density of this process is
\begin{equation*}
    [Z_{t + \Delta} = z_{t+\Delta} \vert Z_t = z_t ] = \phi(z_{t + \Delta};\ z_t + r_t \Delta,\ s_t^2 \Delta),
\end{equation*}
where $\phi(x;\ m,\ v)$ is the pdf of the normal distribution with mean $m$ and variance $v$.

\paragraph{Geometric Brownian motion} The process $Z_t$ is called geometric Brownian motion if $\log(Z_t)$ follows a Brownian motion with drift. The varying-coefficient geometric Brownian motion is a diffusion process with $\mu(Z_t, \bm\theta_t) = r_t Z_t$ and $\sigma(Z_t, \bm\theta_t) = s_t Z_t$. Geometric Brownian motion is a popular choice to model population growth in ecology, where $Z_t$ is the population size at time $t$, and $r_t$ is the growth rate \citep{dennis1991}, which could be modelled as a function of covariates in the framework that we present. 

This process is also used in finance to describe asset prices under the Black-Scholes model. In addition, stochastic volatility models are typically based on geometric Brownian motion, where the variance parameter $s_t^2$ is modelled with an SDE \citep{ait2007}. The varying-coefficient model introduced here is an alternative approach to specify a process with time-varying volatility.

The standard geometric Brownian motion has a closed form transition density, and so we can obtain the transition density of the varying-coefficient process by substituting $r_t$ and $s_t$ for the SDE parameters,
\begin{equation*}
    [Z_{t + \Delta} = z_{t+\Delta} \vert Z_t = z_t ] = 
        \frac{1}{\sqrt{2 \pi \Delta}} \frac{1}{z_{t+\Delta} s_t}
        \exp \left[ - \frac{\left\{ \log(z_{t+\Delta}) - \log(z_t) - (r_t - 0.5 s_t^2) \Delta \right\}^2} 
            {2 s_t^2 \Delta} \right].
\end{equation*}

\begin{figure}[htbp]
  \centering
  \includegraphics[width=\textwidth]{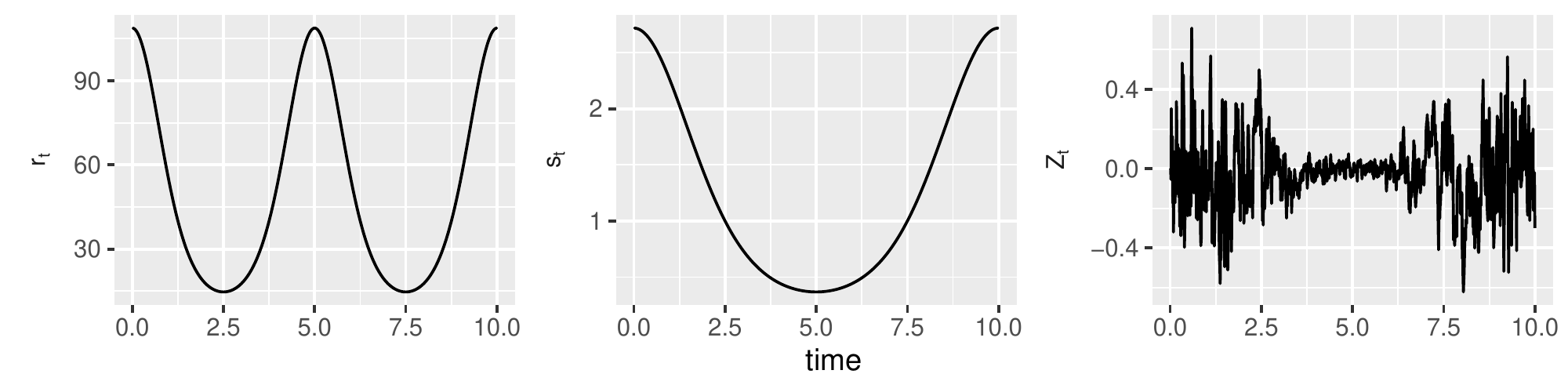}
  \caption{Simulated example for the time-varying Ornstein-Uhlenbeck process with mean $\zeta = 0$, showing the reversion parameter $r_t$ (left), the diffusion parameter $s_t$ (middle) and the simulated process $Z_t$ (right).}
  \label{fig:SDE_example}
\end{figure}

\paragraph{Ornstein-Uhlenbeck (OU)} The varying-coefficient OU process is obtained with $\mu(Z_t, \bm\theta_t) = r_t (\zeta - Z_t)$ and $\sigma(Z_t, \bm\theta_t) = s_t$. It is mean-reverting, i.e.\ it tends to revert to its mean value $\zeta$ with a rate measured by $r_t$ and diffusion measured by $s_t$. The standard OU process was proposed by \cite{uhlenbeck1930} to describe the velocity of a particle subject to friction, and it has also been proposed as a model for interest rates in finance \citep{vasicek1977}. Figure \ref{fig:SDE_example} shows a realisation from the varying-coefficient OU process, to illustrate how time-varying parameters can induce time-varying dynamics. The approximate transition density of the varying-coefficient OU process is
\begin{equation*}
    [Z_{t + \Delta} = z_{t+\Delta} \vert Z_t = z_t ] = \phi \left\{ 
            z_{t + \Delta};\ 
            \zeta + e^{-r_t \Delta} (z_t - \zeta),\ 
            \frac{s_t^2}{2r_t} (1 - e^{-2 r_t \Delta}) 
        \right\}.
\end{equation*}

\cite{dunn1977} suggested the bivariate OU process as a model of movement for animals displaying home range behaviour: in two dimensions, mean reversion is analogous to attraction to a point in space, such as the centre of an animal's home range. The long-term distribution of the OU process is Gaussian, and can be derived in closed form. In that context, $r_t$ and $s_t$ can be linked to the size of the home range, and the strength of attraction to a central location. In practice, $V_t$ and $Z_t$ are usually two-dimensional (representing Easting and Northing), but it is most common to assume that the process is isotropic, i.e.\ the same process describes the movement in both dimensions. \cite{blackwell1997} argued that the OU process may often be too simplistic to model the movement of animals, and proposed a mixture model, where an animal switches between discrete behaviours through time, each associated with an OU process. In the framework presented here, $r_t$ and $s_t$ are specified as flexible functions of covariates, as described in Section 2 of the manuscript, and the dynamics of the OU process can therefore change smoothly over time and space. The number of time-varying parameters in the model is not limited to two ($r_t$ and $s_t$), and the centre of attraction $\zeta$ could in principle also be formulated as a function of covariates.

The OU process is also a popular model for the \emph{velocity} of a moving animal, as proposed by \cite{johnson2008}. In the varying-coefficient variant of this model, the velocity $V_t$ is specified as $dV_t = r_t (\zeta - V_t) dt + s_t dW_t$, and the location $Z_t$ of the animal is obtained as the integral $Z_t = Z_0 + \int_0^t V_s ds$, where $Z_0$ is the initial location. Here, the two parameters $r_t$ and $s_t$ can be linked to the speed and persistence of the movement, and $\zeta$ is usually set to zero to indicate that there is no systematic bias in the velocity. Several approaches have been developed to allow for time-varying dynamics in the velocity OU model. In particular, \cite{michelot2019ctcrw} presented a state-switching formulation, where an animal transitions between different velocity OU processes through time, characterised by different movement characteristics. \cite{russell2018} used a time-varying term for the centre of attraction $\zeta$, to include covariate effects on the direction of movement of the animal. The model presented here generalises that approach to the case where any parameter of the velocity process can be written as a GAM of spatiotemporal (or other) covariates.

\paragraph{Potential-based models} In ecology, diffusion processes have also been used to study the response of animals to their environment. For this purpose, the drift of the process can be specified as a function of the gradient of a ``potential'' function $H$, which measures habitat suitability over space \citep{preisler2004}. In that model, we have $\mu(Z_t, \bm\theta_t) = -\nabla H(Z_t, \bm\theta_t)$, where $\nabla$ is the spatial gradient operator. \cite{preisler2004} estimated $H$ with smoothing splines of spatial covariates, using methodology similar to that presented in this paper. Within the framework that we propose, their model could be extended to include non-spatial covariates, and to investigate covariate effects on the diffusion parameter of the process (which they assumed constant).

\section*{Appendix B\quad Implementation}

\subsection*{B.1\quad Model matrices using mgcv}

The mgcv R package can be used to define the design matrices (including basis functions) and the penalty matrix for basis-penalty smooths, with the function \texttt{gam} \citep{wood2017}. As an example, consider the data frame shown below, where `ID' is the time series identifier, `Z' is the response variable, and `x1' and `x2' are two covariates.

\begin{knitrout}
\definecolor{shadecolor}{rgb}{0.969, 0.969, 0.969}\color{fgcolor}\begin{kframe}
\begin{verbatim}
##   ID           Z       x1         x2
## 1  1 -1.06520105 1.139137 -0.3088276
## 2  1 -1.36064381 2.159093 -0.1593682
## 3  1 -1.04660349 2.610276 -0.6394209
## 4  1  0.35250424 1.830601 -0.7986991
## 5  1 -0.07465397 1.905763  0.4524077
## 6  1 -0.91468379 2.124810  0.3858297
\end{verbatim}
\end{kframe}
\end{knitrout}

Then, consider that \texttt{Z} is to be modelled with a varying-coefficient SDE, where the relationship of each SDE parameter with the covariates is specified by the following formula,
\begin{knitrout}
\definecolor{shadecolor}{rgb}{0.969, 0.969, 0.969}\color{fgcolor}\begin{kframe}
\begin{alltt}
\hlcom{# Formula for SDE parameter, using mgcv syntax for smooth}
\hlcom{# terms and random effects}
\hlstd{form} \hlkwb{<-} \hlopt{~} \hlstd{x1} \hlopt{+} \hlkwd{s}\hlstd{(x2,} \hlkwc{k} \hlstd{=} \hlnum{5}\hlstd{,} \hlkwc{bs} \hlstd{=} \hlstr{"ts"}\hlstd{)} \hlopt{+} \hlkwd{s}\hlstd{(ID,} \hlkwc{bs} \hlstd{=} \hlstr{"re"}\hlstd{)}
\end{alltt}
\end{kframe}
\end{knitrout}
i.e.\ \texttt{x1} has a linear effect, \texttt{x2} has a smooth effect modelled using thin-plate regression splines, and a random normal intercept is included for \texttt{ID}. (See the documentation of mgcv for additional detail on the syntax.) The design matrices can then be derived as follows,
\begin{knitrout}
\definecolor{shadecolor}{rgb}{0.969, 0.969, 0.969}\color{fgcolor}\begin{kframe}
\begin{alltt}
\hlcom{# Create smooth object using mgcv}
\hlstd{smooth} \hlkwb{<-} \hlkwd{gam}\hlstd{(}\hlkwc{formula} \hlstd{=} \hlkwd{update}\hlstd{(form, dummy} \hlopt{~} \hlstd{.),}
              \hlkwc{data} \hlstd{=} \hlkwd{cbind}\hlstd{(}\hlkwc{dummy} \hlstd{=} \hlnum{1}\hlstd{, data),}
              \hlkwc{fit} \hlstd{=} \hlnum{FALSE}\hlstd{)}

\hlcom{# Design matrix}
\hlstd{X} \hlkwb{<-} \hlstd{smooth}\hlopt{$}\hlstd{X}

\hlcom{# Number of non-smooth model terms (i.e. fixed effects)}
\hlstd{nsdf} \hlkwb{<-} \hlstd{smooth}\hlopt{$}\hlstd{nsdf}

\hlcom{# Design matrix for fixed effects}
\hlstd{X_fe} \hlkwb{<-} \hlstd{X[,} \hlnum{1}\hlopt{:}\hlstd{nsdf,} \hlkwc{drop} \hlstd{=} \hlnum{FALSE}\hlstd{]}

\hlcom{# Design matrix for random effects (including smooth model terms)}
\hlstd{X_re} \hlkwb{<-} \hlstd{X[,} \hlopt{-}\hlstd{(}\hlnum{1}\hlopt{:}\hlstd{nsdf),} \hlkwc{drop} \hlstd{=} \hlnum{FALSE}\hlstd{]}
\end{alltt}
\end{kframe}
\end{knitrout}

The design matrix for the fixed effects is
\begin{knitrout}
\definecolor{shadecolor}{rgb}{0.969, 0.969, 0.969}\color{fgcolor}\begin{kframe}
\begin{alltt}
\hlkwd{head}\hlstd{(X_fe)}
\end{alltt}
\begin{verbatim}
##   (Intercept)       x1
## 1           1 1.139137
## 2           1 2.159093
## 3           1 2.610276
## 4           1 1.830601
## 5           1 1.905763
## 6           1 2.124810
\end{verbatim}
\end{kframe}
\end{knitrout}
and the design matrix for the random effects is
\begin{knitrout}
\definecolor{shadecolor}{rgb}{0.969, 0.969, 0.969}\color{fgcolor}\begin{kframe}
\begin{alltt}
\hlkwd{head}\hlstd{(X_re)}
\end{alltt}
\begin{verbatim}
##                                              ID1 ID2 ID3 ID4
## 1 -0.2064242 0.6333093 0.4155266 -0.12492927   1   0   0   0
## 2 -0.1186081 0.6690554 0.4211488 -0.06103063   1   0   0   0
## 3 -0.3930194 0.5225471 0.3865907 -0.26626842   1   0   0   0
## 4 -0.4775210 0.4553437 0.3648461 -0.33436490   1   0   0   0
## 5  0.2464654 0.7122938 0.3947111  0.20052304   1   0   0   0
## 6  0.2071078 0.7157652 0.4013989  0.17205880   1   0   0   0
\end{verbatim}
\end{kframe}
\end{knitrout}
where the first four columns correspond to the four basis functions for \texttt{x2}, and the four last columns are dummy indicator variables for \texttt{ID}. Then, the linear predictor for the SDE parameter is
\begin{knitrout}
\definecolor{shadecolor}{rgb}{0.969, 0.969, 0.969}\color{fgcolor}\begin{kframe}
\begin{alltt}
\hlstd{lp} \hlkwb{<-} \hlstd{X_fe} \hlopt{%*%} \hlstd{coeff_fe} \hlopt{+} \hlstd{X_re} \hlopt{%*%} \hlstd{coeff_re}
\end{alltt}
\end{kframe}
\end{knitrout}
where \texttt{coeff\_fe} and \texttt{coeff\_re} are the coefficients for the fixed effects and the random effects, respectively. The SDE parameter (at each time point) is then obtained by applying the inverse link function to this linear predictor.

Similarly, the penalty matrix for the smooth terms can be extracted from the GAM object,
\begin{knitrout}
\definecolor{shadecolor}{rgb}{0.969, 0.969, 0.969}\color{fgcolor}\begin{kframe}
\begin{alltt}
\hlstd{S} \hlkwb{<-} \hlstd{smooth}\hlopt{$}\hlstd{S}
\end{alltt}
\end{kframe}
\end{knitrout}

\subsection*{B.2\quad Laplace approximation using TMB}

The model fitting procedure proposed in the paper requires the evaluation of the marginal likelihood, where the random effects (including basis coefficients for smooth terms) have been integrated out. We suggest using the R package Template Model Builder (TMB) to implement the marginal likelihood, based on the Laplace approximation \cite{kristensen2016}. Here, we broadly explain how TMB can be used to evaluate the objective function (i.e., the negative log-likelihood) of the model, and to obtain point and uncertainty estimates for the model parameters.

The joint log-likelihood must first be written in C++, following the TMB syntax (for examples, see \url{kaskr.github.io/adcomp/examples.html}). For our model, it has two main components:
\begin{enumerate}
    \item the joint log-likelihood of the fixed and random effect parameters, given by the sum of the log-pdf of the observed transitions, $\log [Z_{i+1} \vert Z_i]$, e.g.\ obtained using the Euler-Maruyama discretization of the process. This part requires the SDE parameters on a time grid, which can be computed based on the model matrices provided by mgcv, as described in Appendix B.1.
    \item the log-pdf of the basis coefficients (and other random effects) given the smoothness parameter, $\log [\bm\beta \vert \bm\lambda]$, obtained as the log of a multivariate normal pdf with block-diagonal precision matrix, where the $i$-th block is $\lambda_i \bm{S}_i$. The smoothness matrix $\bm{S}_i$ is provided by mgcv, as described in Appendix B.1.
\end{enumerate}

The C++ function takes the following arguments as data:
\begin{itemize}
\item times of observations $(t_1, \dots, t_n)$;
\item observations $(z_1, \dots, z_n)$
\item design matrix for fixed effects (\texttt{X\_fe} in Appendix B.1), provided by mgcv;
\item design matrix for random effects (\texttt{X\_re} in Appendix B.1), provided by mgcv;
\item smoothness matrix (\texttt{S} in Appendix B.1), provided by mgcv;
\end{itemize}
and the following arguments as parameters:
\begin{itemize}
\item coefficients for fixed effects (\texttt{coeff\_fe} in Appendix B.1);
\item coefficients for random effects (\texttt{coeff\_re} in Appendix B.1);
\item smoothness parameters $\bm\lambda$.
\end{itemize}

The objective function (and its gradient) can then be defined in R with the TMB function \texttt{MakeADFun} applied to the joint negative log-likelihood, using the argument `random' to specify that the basis coefficients \texttt{coeff\_re} should be treated as random effects. It can then be passed to a numerical optimiser, e.g.\ \texttt{optim} or \texttt{nlm}, to perform maximum likelihood estimation of the fixed effect parameters. After optimisation, the function \texttt{sdreport} can be used to obtain estimates of the random effect parameters, as well as a joint precision matrix for the fixed and random effects. Posterior samples of all model parameters can be generated from a multivariate normal distribution, where the covariance matrix is the inverse of this precision matrix.

\subsection*{B.3\quad Package smoothSDE}

This method is implemented in the R package \texttt{smoothSDE}, available at \url{github.com/TheoMichelot/smoothSDE}. The package provides functions for model fitting, uncertainty estimation, model checking, and model plots, and we hope that it will greatly facilitate the application of varying-coefficient SDEs. Here, we present the code required to fit the varying-coefficient model applied to the elephant data set from \cite{wall2014}, in Section 3.2 of the main text, to showcase its use.

First, we download the data from the Movebank data repository, and create a data frame with columns for ID, time (as numeric), response variables (here, easting and northing), and covariates (here, temperature),
\begin{knitrout}
\definecolor{shadecolor}{rgb}{0.969, 0.969, 0.969}\color{fgcolor}\begin{kframe}
\begin{alltt}
\hlcom{# Load package and set seed for reproducibility}
\hlkwd{library}\hlstd{(smoothSDE)}
\hlkwd{set.seed}\hlstd{(}\hlnum{58652}\hlstd{)}

\hlcom{# Load data and keep relevant columns}
\hlstd{URL} \hlkwb{<-} \hlkwd{paste0}\hlstd{(}\hlstr{"https://www.datarepository.movebank.org/bitstream/handle/"}\hlstd{,}
              \hlstr{"10255/move.373/Elliptical%20Time-Density%20Model%20%28Wall%"}\hlstd{,}
              \hlstr{"20et%20al.%202014%29%20African%20Elephant%20Dataset%20%"}\hlstd{,}
              \hlstr{"28Source-Save%20the%20Elephants%29.csv"}\hlstd{)}
\hlstd{raw} \hlkwb{<-} \hlkwd{read.csv}\hlstd{(}\hlkwd{url}\hlstd{(URL))}
\hlstd{keep_cols} \hlkwb{<-} \hlkwd{c}\hlstd{(}\hlnum{11}\hlstd{,} \hlnum{13}\hlstd{,} \hlnum{14}\hlstd{,} \hlnum{17}\hlstd{,} \hlnum{6}\hlstd{)}
\hlstd{raw_cols} \hlkwb{<-} \hlstd{raw[, keep_cols]}
\hlkwd{colnames}\hlstd{(raw_cols)} \hlkwb{<-} \hlkwd{c}\hlstd{(}\hlstr{"ID"}\hlstd{,} \hlstr{"x"}\hlstd{,} \hlstr{"y"}\hlstd{,} \hlstr{"date"}\hlstd{,} \hlstr{"temp"}\hlstd{)}

\hlcom{# Only keep five months to eliminate seasonal effects}
\hlstd{track} \hlkwb{<-} \hlkwd{subset}\hlstd{(raw_cols, ID} \hlopt{==} \hlkwd{unique}\hlstd{(ID)[}\hlnum{1}\hlstd{])}
\hlstd{track}\hlopt{$}\hlstd{date} \hlkwb{<-} \hlkwd{as.POSIXlt}\hlstd{(track}\hlopt{$}\hlstd{date,} \hlkwc{tz} \hlstd{=} \hlstr{"GMT"}\hlstd{)}
\hlstd{track}\hlopt{$}\hlstd{time} \hlkwb{<-} \hlkwd{as.numeric}\hlstd{(track}\hlopt{$}\hlstd{date} \hlopt{-} \hlkwd{min}\hlstd{(track}\hlopt{$}\hlstd{date))}\hlopt{/}\hlnum{3600}
\hlstd{keep_rows} \hlkwb{<-} \hlkwd{which}\hlstd{(track}\hlopt{$}\hlstd{date} \hlopt{>} \hlkwd{as.POSIXct}\hlstd{(}\hlstr{"2009-05-01 00:00:00"}\hlstd{)} \hlopt{&}
                       \hlstd{track}\hlopt{$}\hlstd{date} \hlopt{<} \hlkwd{as.POSIXct}\hlstd{(}\hlstr{"2009-09-30 23:59:59"}\hlstd{))}
\hlstd{track} \hlkwb{<-} \hlstd{track[keep_rows,]}

\hlcom{# Convert to km}
\hlstd{track}\hlopt{$}\hlstd{x} \hlkwb{<-} \hlstd{track}\hlopt{$}\hlstd{x}\hlopt{/}\hlnum{1000}
\hlstd{track}\hlopt{$}\hlstd{y} \hlkwb{<-} \hlstd{track}\hlopt{$}\hlstd{y}\hlopt{/}\hlnum{1000}
\end{alltt}
\end{kframe}
\end{knitrout}

\begin{knitrout}
\definecolor{shadecolor}{rgb}{0.969, 0.969, 0.969}\color{fgcolor}\begin{kframe}
\begin{alltt}
\hlcom{# Data set including ID, time, responses (x, y), and covariate (temp)}
\hlkwd{head}\hlstd{(track)}
\end{alltt}
\begin{verbatim}
##               ID        x        y                date temp time
## 9689 Salif Keita 572.3427 1675.424 2009-05-01 00:00:00   33 9703
## 9690 Salif Keita 572.5443 1675.392 2009-05-01 01:00:00   32 9704
## 9691 Salif Keita 572.6159 1675.339 2009-05-01 02:00:00   31 9705
## 9692 Salif Keita 572.7745 1675.101 2009-05-01 03:00:00   31 9706
## 9693 Salif Keita 572.8844 1675.065 2009-05-01 04:00:00   31 9707
## 9694 Salif Keita 573.6659 1674.322 2009-05-01 05:00:00   30 9708
\end{verbatim}
\end{kframe}
\end{knitrout}

For this analysis, we use the velocity Ornstein-Uhlenbeck model (also called continuous-time correlated random walk, ``CTCRW''), which has two parameters: `beta' (mean reversion parameter) and `sigma' (variance parameter), defined in \cite{johnson2008}. We define formulas for the SDE parameters to express covariate dependence, using the syntax from mgcv for smooth terms and random effects. Both parameters are specified as functions of the temperature covariate, using thin-plate regression splines, 
\begin{knitrout}
\definecolor{shadecolor}{rgb}{0.969, 0.969, 0.969}\color{fgcolor}\begin{kframe}
\begin{alltt}
\hlcom{# Model formulas for CTCRW parameters}
\hlstd{formulas} \hlkwb{<-} \hlkwd{list}\hlstd{(}\hlkwc{beta} \hlstd{=} \hlopt{~} \hlkwd{s}\hlstd{(temp,} \hlkwc{k} \hlstd{=} \hlnum{10}\hlstd{,} \hlkwc{bs} \hlstd{=} \hlstr{"ts"}\hlstd{),}
                 \hlkwc{sigma} \hlstd{=} \hlopt{~} \hlkwd{s}\hlstd{(temp,} \hlkwc{k} \hlstd{=} \hlnum{10}\hlstd{,} \hlkwc{bs} \hlstd{=} \hlstr{"ts"}\hlstd{))}
\end{alltt}
\end{kframe}
\end{knitrout}

We then create the SDE as an object, which encapsulates the data and model formulas,
\begin{knitrout}
\definecolor{shadecolor}{rgb}{0.969, 0.969, 0.969}\color{fgcolor}\begin{kframe}
\begin{alltt}
\hlcom{# Type of SDE model: continuous-time correlated random walk}
\hlstd{type} \hlkwb{<-} \hlstr{"CTCRW"}

\hlcom{# Create SDE model object}
\hlstd{my_sde} \hlkwb{<-} \hlstd{SDE}\hlopt{$}\hlkwd{new}\hlstd{(}\hlkwc{formulas} \hlstd{= formulas,} \hlkwc{data} \hlstd{= track,} \hlkwc{type} \hlstd{= type,}
                  \hlkwc{response} \hlstd{=} \hlkwd{c}\hlstd{(}\hlstr{"x"}\hlstd{,} \hlstr{"y"}\hlstd{))}

\hlcom{# Fit model}
\hlstd{my_sde}\hlopt{$}\hlkwd{fit}\hlstd{()}
\end{alltt}
\end{kframe}

\end{knitrout}

After model fitting, estimates of the SDE parameters, and posterior samples, can be plotted as functions of the covariates,
\begin{knitrout}
\definecolor{shadecolor}{rgb}{0.969, 0.969, 0.969}\color{fgcolor}\begin{kframe}
\begin{alltt}
\hlstd{my_sde}\hlopt{$}\hlkwd{plot_par}\hlstd{(}\hlstr{"temp"}\hlstd{,} \hlkwc{n_post} \hlstd{=} \hlnum{100}\hlstd{)}
\end{alltt}
\end{kframe}
\includegraphics[width=\linewidth]{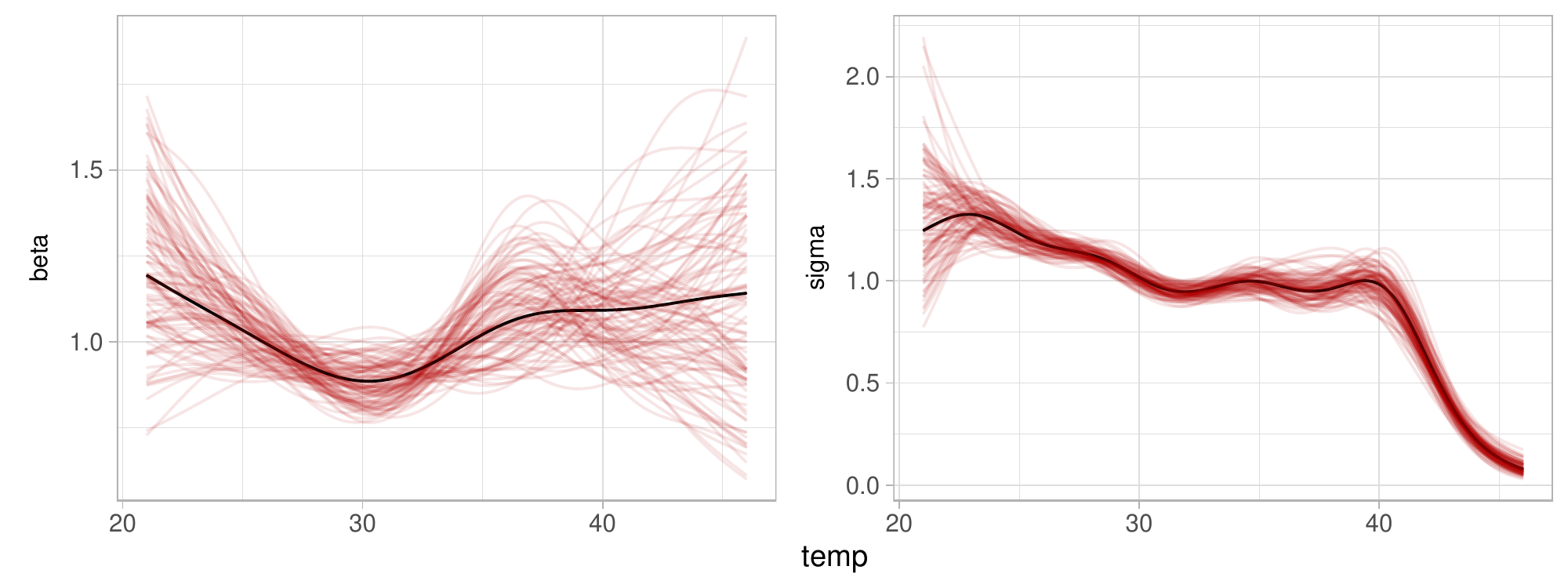} 

\end{knitrout}

\section*{Appendix C\quad Simulation study}

We ran simulations for two different model formulations, to investigate the performance of the inference method presented in Section 3 of the manuscript.

\begin{figure}[htbp]
  \centering
  \includegraphics[width=\textwidth]{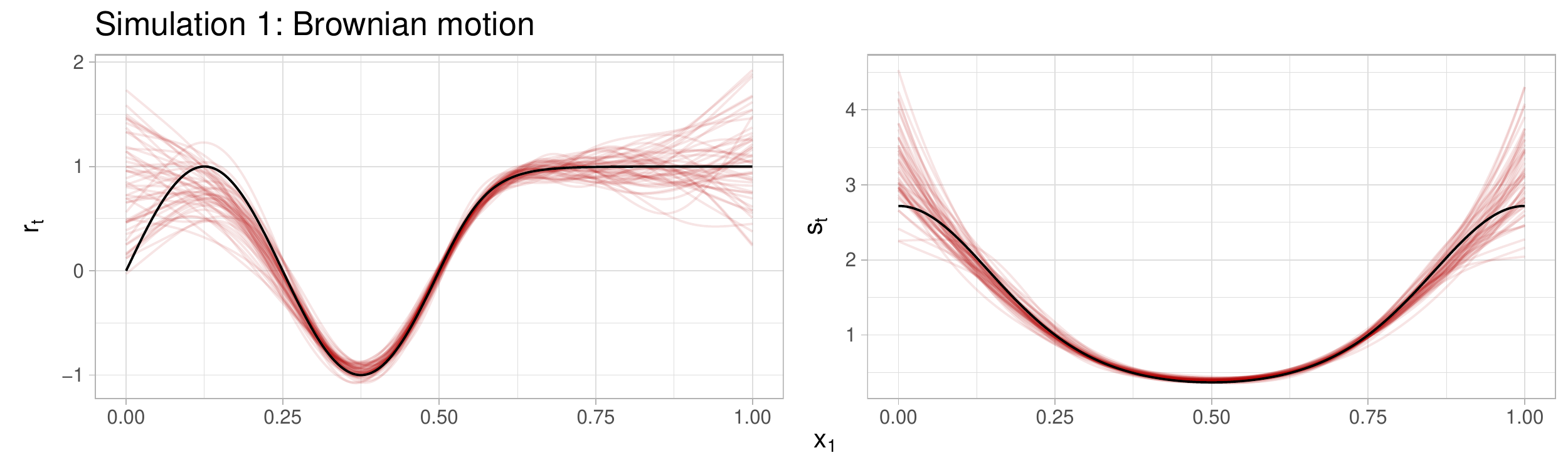} \\
  \includegraphics[width=\textwidth]{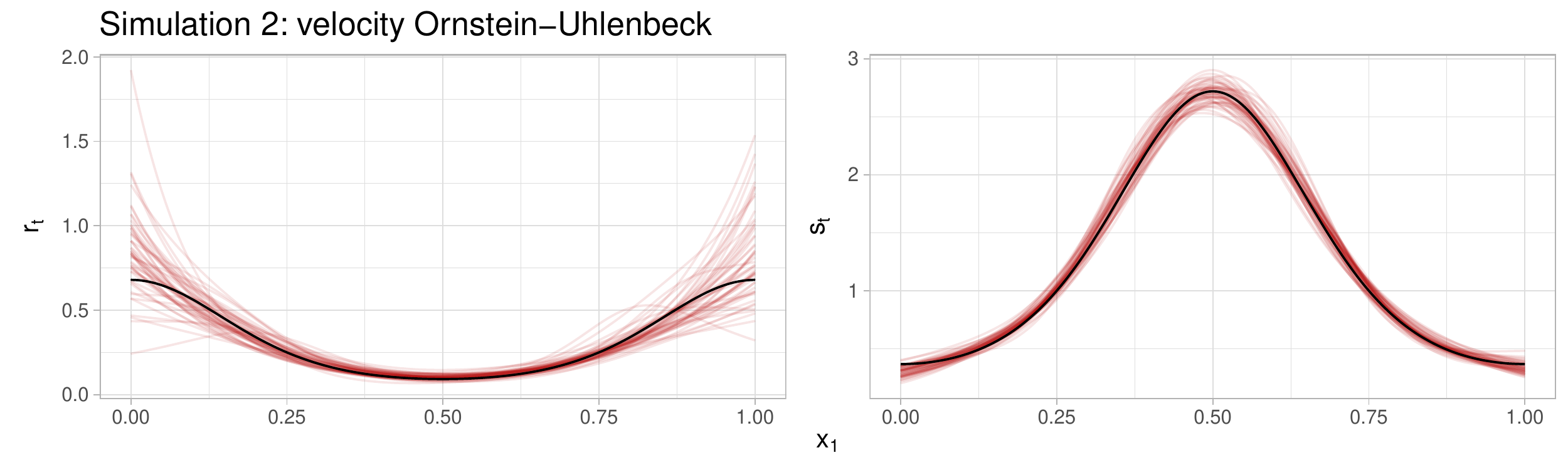}
  \caption{Results of simulation scenario 1 (top row) and scenario 2 (bottom row). The red lines are the 50 estimated drift parameters $r_t$ and diffusion parameters $s_t$ as functions of the covariate $x_1$, and the black lines are the true functions.}
  \label{fig:sim_res}
\end{figure}

\paragraph{Scenario 1} We considered a Brownian motion with drift, where the parameters $r_t$ (drift) and $s_t$ (diffusion) were functions of a covariate $x_{1t}$. We generated the covariate as Brownian motion (with zero drift) over the time period of the simulations, and then scaled it to $[0, 1]$. For a choice of functions $r_t$ and $s_t$ (shown in Figure \ref{fig:sim_res}), we ran 50 simulations. For each, we simulated $10^5$ data points at a fine time resolution ($\Delta = 0.01$) to make sure that the discretization error of simulation was negligible. We then downsampled the time series by keeping 2000 observations at random, to assess the performance of the method to recover the model parameters from data collected at irregular time intervals. We fitted the model to each downsampled data set, and derived estimates of $r_t$ and $s_t$. Figure \ref{fig:sim_res} shows the true $r_t$ and $s_t$ and the 50 estimates as functions of the covariate $x_{1t}$. The splines generally fitted the true functions well, although most of them did not capture the detailed oscillations in the drift parameter $r_t$ over the lower end of the covariate range. This is because the smoothness of the true function varied over the covariate range (less smooth for low values, more smooth for high values), and the estimated smoothness can therefore be viewed as an average. This could be addressed with adaptive smoothing, i.e.\ by letting the degree of smoothness vary across values of the covariate \citep[][Section 5.3]{wood2017}.

\paragraph{Scenario 2} In the second scenario, we assessed the performance of maximum likelihood estimation in the case where the diffusion process is not directly observed, and the Kalman filter needs to be implemented, as described in Section 2.2.1. We simulated 50 trajectories from the velocity OU model presented in Appendix A, following the same procedure as in the first scenario to obtain irregular observations. The parameters $r_t$ and $s_t$ used in the simulations were functions of one covariate, like in the previous scenario. We implemented the Kalman filter, to make inference about the parameters of the latent velocity process, based on the simulated observations. Figure \ref{fig:sim_res} shows the results from the 50 simulations. The smoothness and general shape of both parameters $r_t$ and $s_t$ were recovered in all experiments.

\paragraph{Confidence interval coverage} We ran another set of simulations to assess the coverage of Wald confidence intervals obtained using the joint precision matrix given by TMB (using \texttt{sdreport} as described in Appendix B.2). We simulated data from a varying-coefficient Brownian motion with drift, using the same drift and diffusion functions as in Scenario 1. Then, we fitted the model using TMB, and generated 1000 posterior samples for $r_t$ and $s_t$ on a grid over the covariate range. We derived pointwise 95\% confidence intervals on that grid, and checked whether they included the true value. We repeated this experiment 1000 times, and derived the proportions of confidence intervals that included the true value. Average coverage over the covariate range was 95.3\% for $r_t$ and 96.7\% for $s_t$, close to the expected 95\%.

\section*{Appendix D\quad Application to oil prices}

This application involves the analysis of a time series of oil prices, inspired by \cite{garcia2017}. We downloaded daily prices on WTI crude oil between 2 January 1986 and 30 March 2020 from the US Energy Information Administration website (\url{eia.gov/dnav/pet/pet_pri_spt_s1_d.htm}, accessed on 31 March 2020), and derived the log-returns as $R_i = \log(P_{i+1}/P_i)$ for day $i$, where $P_i$ is the price of a barrel of crude oil. There were occasional missing values in this daily time series (easily accommodated in this continuous-time framework), resulting in a total of 8628 data points. We then fitted a varying-coefficinet Brownian motion with drift to the log-returns $R_i$, where the drift and diffusion coefficients were specified as functions of the process value $R_i$. Model fitting took around 1 min on a 1.3GHz Intel i7 CPU. 

\begin{figure}[htbp]
  \centering
  \includegraphics[width=\textwidth]{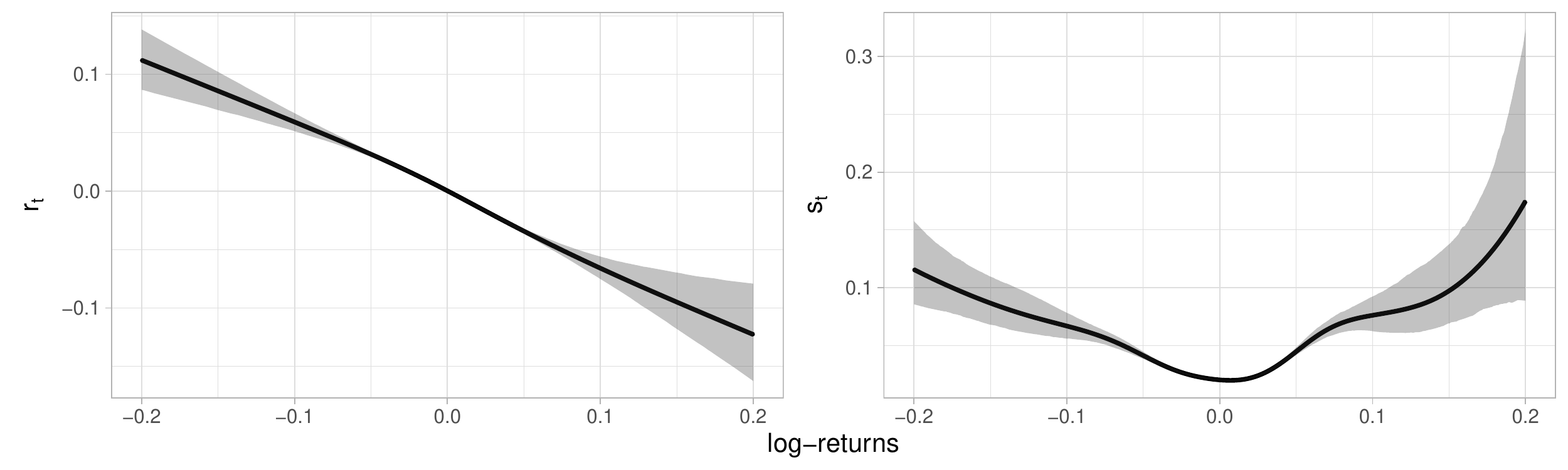}
  \caption{Results of crude oil price analysis, showing the drift parameter $r_t$ and diffusion parameter $s_t$ as functions of the log-returns. The black lines are the mean estimates, and the grey shaded areas are 95\% confidence bands.}
  \label{fig:oil_res}
\end{figure}

Estimates of the drift parameter $r_t$ and diffusion parameter $s_t$ are shown as functions of the log-returns in Figure \ref{fig:oil_res}. Similarly to \cite{garcia2017}, we found that the drift $r_t$ was negative for positive log-returns, and positive for negative log-returns, i.e., the process was attracted to zero. The diffusion parameter $s_t$ increased with the absolute value of the log-returns, meaning that the process was more volatile when it took large (positive or negative) values. Based on a similar observation, \cite{garcia2017} suggested that $s_t$ may be expressed as a quadratic function of the log-returns. Interestingly, our results indicate that this may be inappropriate, because the rate of increase of $s_t$ decreases below $-0.07$ and above $0.07$.

\section*{Appendix E\quad Supplementary information for beaked whale analysis}

\subsection*{E.1\quad Euler angles}

\begin{figure}[htbp]
  \centering
  \includegraphics[width=0.5\textwidth]{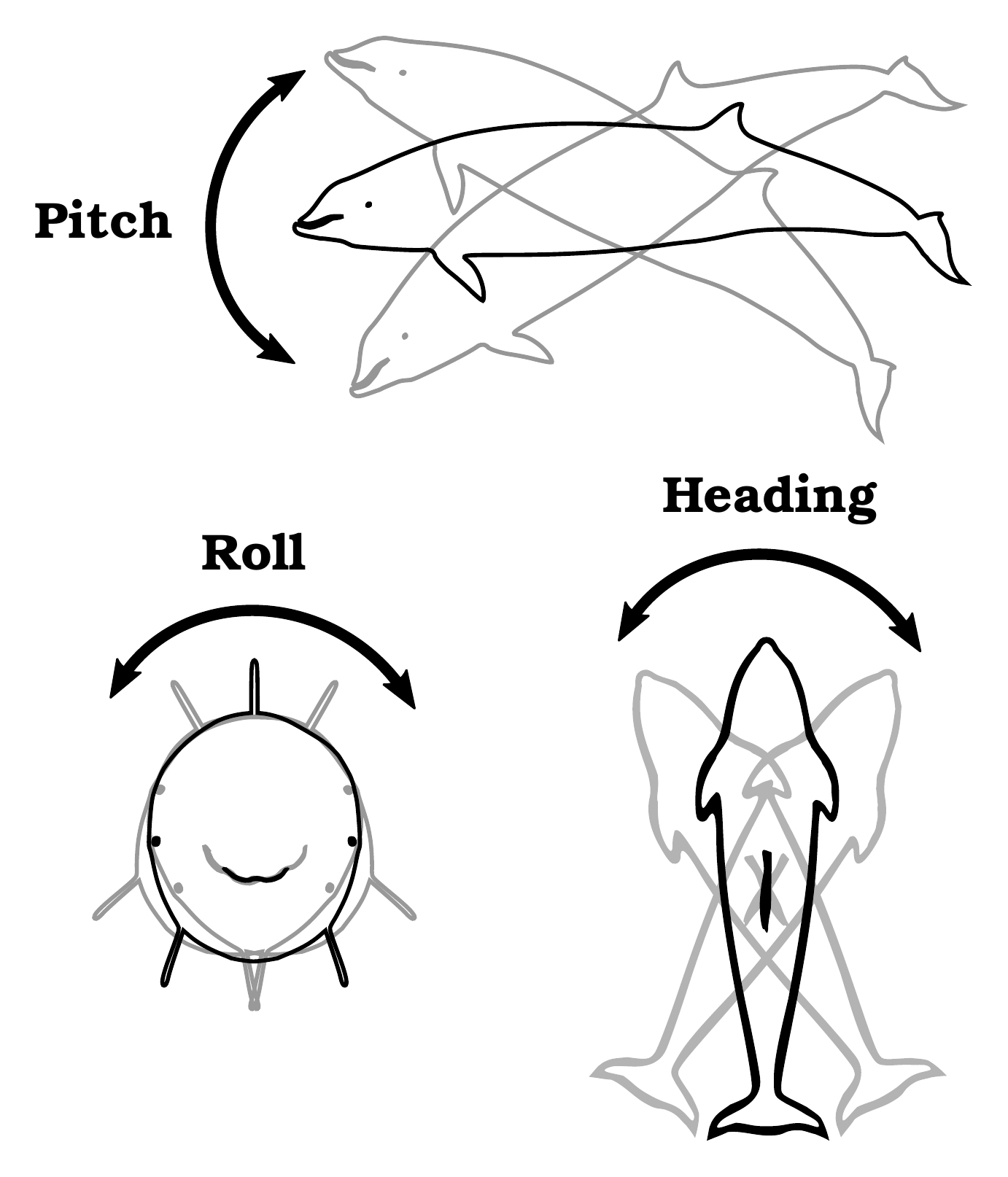}
  \caption{Illustration of Euler angles.}
  \label{fig:euler}
\end{figure}

Figure \ref{fig:euler} shows an illustration of the three Euler angles derived from accelerometer data in the beaked whale analysis of Section 3.3: pitch, roll, and heading. They quantify the posture of the whale in the water.

\subsection*{E.2\quad Heavy-tailed model formulation}

The three variables used in the beaked whale analysis of Section 3.3 of the main paper (pitch, roll, and heading) had heavy-tailed increments, which did not satisfy the assumption of normality made by Brownian motion. Here, we describe how we modified that model to accommodate the heavy tails.

\paragraph{Formulation}
Let $D_i = Z_{t_{i+1}} - Z_{t_i}$ denote the increment in the process between two successive times of observation $t_i$ and $t_{i+1}$. We write each process increment as
\begin{equation*}
  D_i = r_{t_i} \Delta_i + \tilde{s}_{t_i} \sqrt{\Delta_i} X_i
\end{equation*}
where
\begin{itemize}
\item $\Delta_i  = t_{i+1} - t_i$ is the time interval;
\item $r_t$ is a time-varying location parameter;
\item $\tilde{s}_t$ is a time-varying scale parameter;
\item for all $i$, $X_i$ follows a Student's t distribution with $\nu > 2$ degrees of freedom.
\end{itemize}

Then, increments of the process follow a heavy-tailed generalized Student's t distribution, with mean $r_{t_i} \Delta_i$ and standard deviation $\tilde{s}_{t_i} \sqrt{\Delta_i} \sqrt{\nu/(\nu-2)}$. This is based on the standard assumption of Brownian motion that the increment mean scales linearly with the time interval, and the increment standard deviation scales linearly with the square root of the time interval. For ease of interpretation, in the paper we present results for the standard deviation parameter $s_t = \tilde{s}_t \sqrt{\nu/(\nu-2)}$ rather than for the scale parameter $\tilde{s}_t$.

\paragraph{Likelihood}
We consider $n$ observations $(z_1, \dots, z_n)$ from the process, collected at times $t_1 < \dots < t_n$. Under the assumption that increments are independent, we can write the full approximate likelihood as
\begin{equation*}
  L(\theta \vert z_1, \dots, z_n) = \prod_{i=2}^n [Z_{t_i} = z_i \vert Z_{t_{i-1}} = z_{i-1}, r_i, \tilde{s}_i],
\end{equation*}
where $r_i = r_{t_i}$ and $\tilde{s}_i = \tilde{s}_{t_i}$.

The transition density, i.e., the pdf of an increment of the process, is given by
\begin{equation*}
  [Z_{t_i} = z_i \vert Z_{t_{i-1}} = z_{i-1}, r_i, \tilde{s}_i] = f_X \left( \frac{z_{i+1} - z_i - r_i \Delta_i}{\tilde{s}_i \sqrt\Delta_i} \right) \times \frac{1}{\tilde{s}_i \sqrt{\Delta_i}},
\end{equation*}
where $f_X$ is the pdf of a t distribution with $\nu$ degrees of freedom, and the term $1/(\tilde{s}_i \sqrt{\Delta_i})$ is the Jacobian of the transformation from $X_i$ to the increment $D_i$.

\paragraph{Residuals}
We define the residuals of this model as
\begin{equation*}
  \epsilon_i = \frac{z_{i+1} - z_i - r_i \Delta_i}{\tilde{s}_i \sqrt\Delta_i}
\end{equation*}
using the notation from the previous sections. Under the assumptions of the model and the discretization, these residuals follow a t distribution with $\nu$ degrees of freedom.

\subsection*{E.3\quad ACF plots of residuals}

\begin{figure}[htbp]
  \centering
  \includegraphics[width=0.32\textwidth]{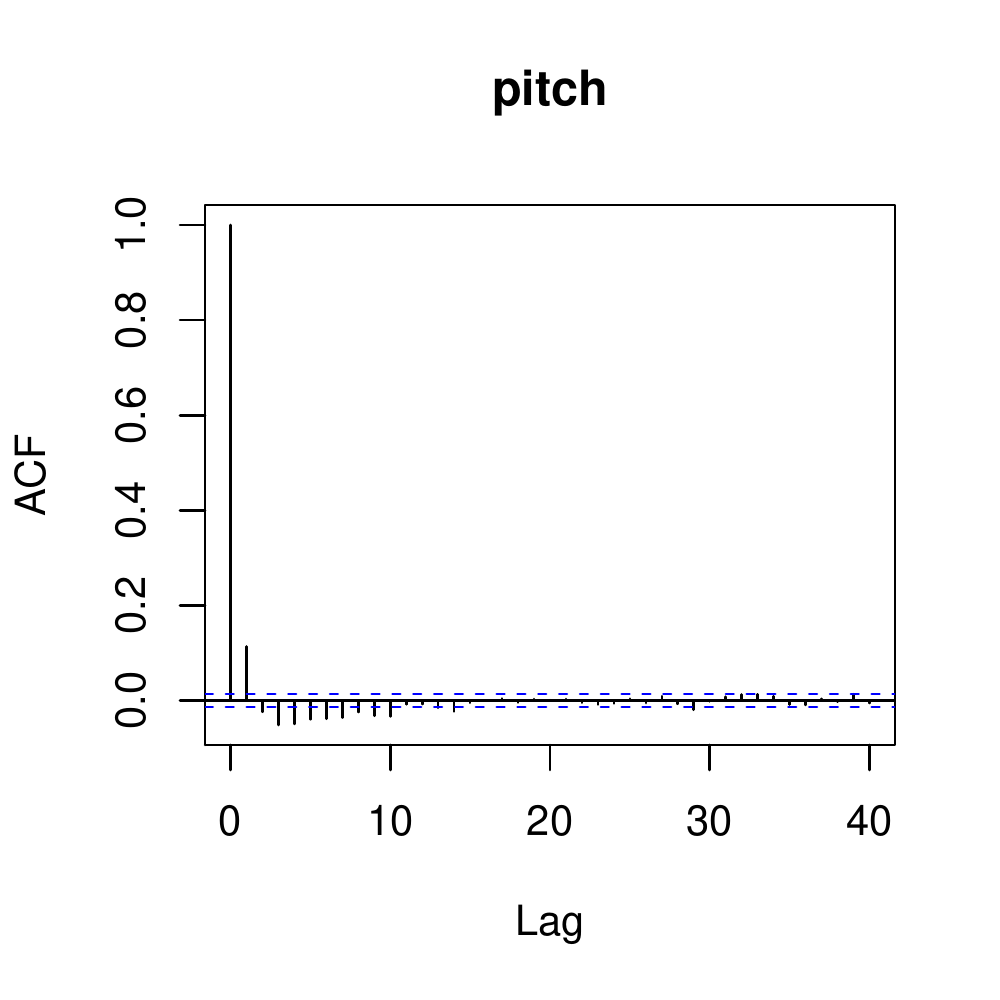}
  \includegraphics[width=0.32\textwidth]{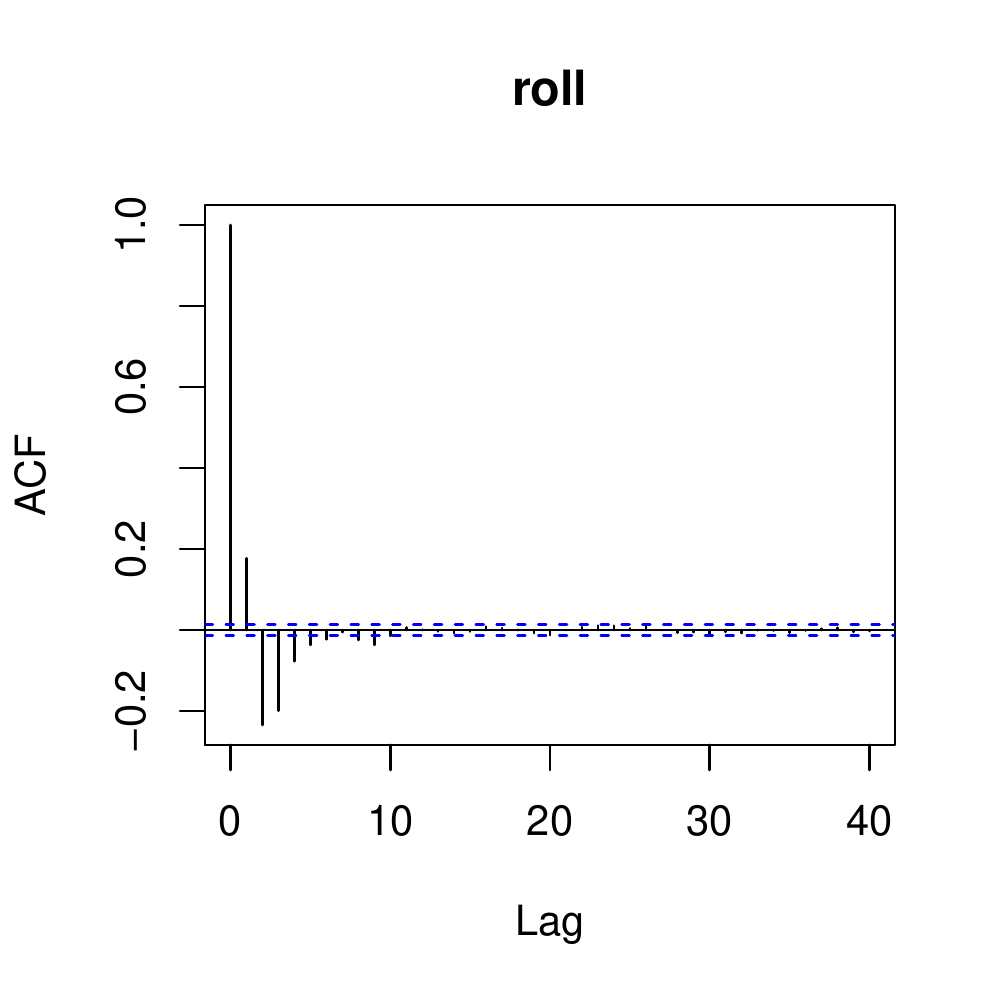}
  \includegraphics[width=0.32\textwidth]{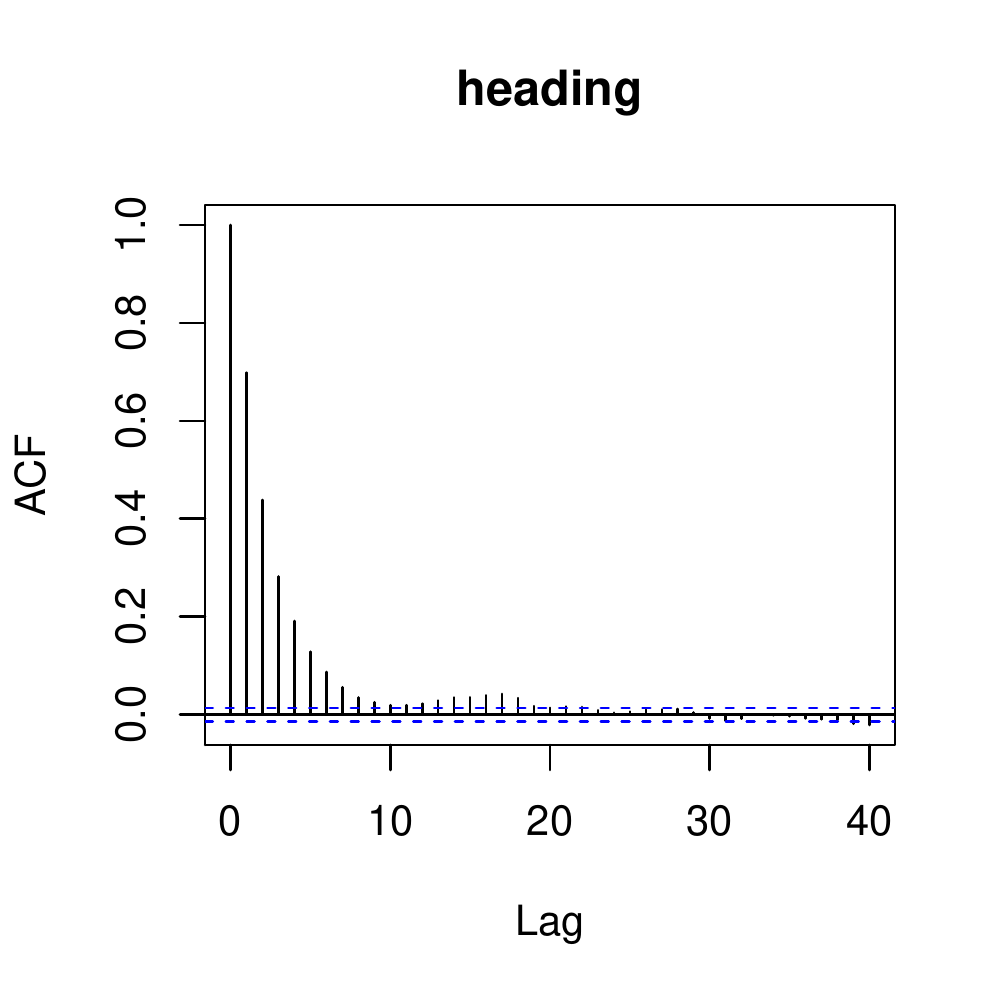}
  \caption{Autocorrelation function plots of residuals for beaked whale analysis. One lag unit corresponds to five seconds.}
  \label{fig:acf}
\end{figure}

Figure \ref{fig:acf} shows autocorrelation function plots of the residuals for the three data variables in the beaked whale analysis (pitch, roll, and heading). Pitch and roll both display some negative autocorrelation over a few time lags (20-30 sec), which may be due to cycles in the motions of whales that were not captured by the model. The high positive autocorrelation in the residuals for heading suggest a strong tendency for whales to persist in the direction of rotation in the horizontal plane (i.e., either circling to the left or to the right).

\end{document}